%% file: manuscript.tex
\documentclass[fleqn,usenatbib]{mnras}

\usepackage[T1]{fontenc}

\DeclareRobustCommand{\VAN}[3]{#2}
\let\VANthebibliography\thebibliography
\def\thebibliography{\DeclareRobustCommand{\VAN}[3]{##3}\VANthebibliography}

\usepackage{graphicx}	
\usepackage{amsmath}	
\usepackage{amssymb}	
\usepackage[nice]{nicefrac}
\usepackage{booktabs}
\usepackage{tabu}
\usepackage{siunitx}
\usepackage{makecell}
\usepackage{multirow}
\usepackage{float}
\usepackage{color}

\newfont{\mlx}{cmssdc10 scaled 770}
\definecolor{pblue}{rgb}{0.0,0.49,0.6745}

\let\oldAA\AA
\renewcommand{\AA}{\text{\normalfont\oldAA}}

\usepackage{newtxtext,newtxmath}
\title[CGCG007-025 chemical composition]{The resolved chemical composition of the starburst dwarf galaxy CGCG007-025: Direct method versus photoionization model fitting.}

\author[V. Fern\'andez]{
V. Fern\'andez$^{1,2}$\thanks{E-mail: vital.fernandez@userena.cl},
R. Amor\'in$^{1,2}$,
R. Sanchez-Janssen$^{3}$,
M. G. del Valle-Espinosa$^{4}$,
P. Papaderos$^{5,6}$
\\
$^{1}$Departamento de Astronom\'ia, Universidad de La Serena, Av. Juan Cisternas 1200 Norte, La Serena, Chile\\
$^{2}$Instituto de Investigaci\'on Multidisciplinar en Ciencia y Tecnolog\'ia, Universidad de La Serena, Ra\'ul Bitr\'an 1305, La Serena, Chile\\
$^{3}$UK Astronomy Technology Centre, Royal Observatory, Blackford Hill, Edinburgh, EH9 3HJ, UK\\
$^{4}$Institute for Astronomy, The University of Edinburgh, Royal Observatory, Blackford Hill, Edinburgh, EH9 3HJ, UK
$^{5}$Centro de Astrof\'isica e Ci\^{e}ncias do Espaço, Universidade de Lisboa - OAL, Tapada da Ajuda, PT1349-018 Lisboa, Portugal\\
$^{6}$Instituto de Astrof\'{i}sica e Ci\^{e}ncias do Espaço - Centro de Astrof\'isica da Universidade do Porto, Rua das Estrelas, 4150-762 Porto, Portugal
}

\date{Accepted XXX. Received YYY; in original form ZZZ}

\pubyear{2015}

\begin{document}
\label{firstpage}
\pagerange{\pageref{firstpage}--\pageref{lastpage}}
\maketitle

\begin{abstract}
This work focuses on the gas chemical composition of CGCG007-025. This compact dwarf is undergoing a galaxy wide star forming burst, whose spatial behaviour has been observed by VLT/MUSE. We present a new line measurement library to treat almost 7800 voxels. The direct method chemical analysis is limited to 484 voxels with good detection of the $[SIII]6312\AA$ temperature diagnostic line. The recombination fluxes are corrected for stellar absorption via a population synthesis. Additionally, we discuss a new algorithm to fit photoionization models via neural networks. The 8 ionic abundances analyzed show a spatial normal distribution with a $\sigma\sim0.1\,dex$, where only half this value can be explained by the uncertainty in the measurements. The oxygen abundance distribution is $12+log(O/H)=7.88\pm0.11$. The $T_{e}[SIII]$ and $ne[SII]$ are also normally distributed. However, in the central and brightest region, the $ne[SII]$ is almost thrice the mean galaxy value. This is also reflected in the extinction measurements. The ionization parameter has a distribution of $log(U) = -2.52^{0.17}_{0.19}$. The parameter spatial behaviour agrees with the $S^{2+}/S^{+}$ map. Finally, the discrepancies between the direct method and the photoionization model fitting are discussed. In the latter technique, we find that mixing lines with uneven uncertainty magnitudes can impact the accuracy of the results. In these fittings, we recommend overestimating the minimum flux uncertainty one order below the maximum line flux uncertainty. This provides a better match with the direct method.

\end{abstract}



\section{Introduction}

The astronomical community will remember 2022 for the successful deployment of the James Webb Space Telescope \citep[see][]{gardner_james_2006}. Among the
many breakthroughs achieved, the JWST early science observations of the SMACS J0723.3-7327 galaxy cluster produced over 30 publications in less than two months\footnote{ADS search including the body "SMACS 0723" and "JWST" keywords}. Focusing on the data from the Near InfraRed Spectrograph (NIRSpec), five emission line galaxies were detected at $5.2 < z < 8.5$, \citep[see][]{carnall_first_2022,schaerer_first_2022,trump_physical_2022}. Despite the fact that this instrument flux calibration has yet to be evaluated \citep[see][]{birkmann_-flight_2022} numerous research groups are reaching similar conclusions: 1) The line flux ratios are in agreement with those observed in Extreme Emission Line Galaxies (EELG). This can be appreciated in the graphical comparisons by \citet{trump_physical_2022} and \citet{brinchmann_high-z_2022} with the SDSS samples of EELGs by \citet{perez-montero_extreme_2021} and \citet{izotov_multi-wavelength_2014} respectively. 2) The metallicities, derived via the direct method, thanks to the $[OIII]4363\text{Å}$
auroral line detection, are around $\approx0.1Z_{\odot}$ \citep[see][]{arellano-cordova_first_2022,curti_chemical_2022}. 3) The
stellar mass tracers predict values within the $7.5\lesssim log(M_{*})\lesssim9.0$ range \citep[see][]{tacchella_jwst_2022}. These measurements lead to the following conclusion: There are still galaxies in the local universe with nebular properties closely resembling those in the 1 Gyr old universe.

This was already anticipated by, e.g., \citet{amorin_oxygen_2010,schaerer_ionizing_2016}, while analysing the chemical content of Green Peas galaxies, in the local and (pre-JWST) high redshift universe \citep[see][]{amorin_star_2012,schaerer_ionizing_2016,amorin_analogues_2017}. Despite their compact sizes $(r_{e}<5kpc)$ and low masses $(8.5<log(M_{*})<10)$ \citep[see][]{cardamone_galaxy_2009}, these galaxies can be easily detected in spectroscopic surveys due to their intense nebular emission.

Moreover, since these emissions are likely constrained to a single filter, their vivid colors highlight them in photometric surveys. Beyond, the radiation from the stellar atmospheres, the main phenomenon dominating the emission properties is the gas composition. Indeed, at the optical range, the cooling of a low metallicity gas produces an enormous photons flux, from the forbidden transitions. Consequently, studying the chemistry of EELG not only improves our understanding on the local star formation phenomena, it can help us interpret the observations from the young universe.

In this manuscript, we present the initial results from an exhaustive analysis of the CGCG007-025 and UGC5205 galaxy pair. As mentioned by \citet{van_zee_evolutionary_2001}, this isolated galaxy pair is very interesting. The authors stated that while the projected separation is $8.3\,kpc$ (at a velocity difference of $84\,km/s$) there is a stark contrast between them: CGCG007-025 is undergoing a massive star formation phase, while UGC5205 is gas depleted. The observation of a tidal tail in UGC5205 encourages the idea that galactic interaction caused this state. CGCG007-025 is included in the COS Legacy Archive Spectroscopy Survey (CLASSY) \citep[see][]{berg_cos_2022} for its high redshift analogue characteristics. Finally, this system proximity, $z\approx0.047$, make it an ideal laboratory to study the spatial properties of star formation enhancing and quenching. 

The current results focus on the gas chemical composition of CGCG007-025, \emph{Catalogue of Galaxies and of Clusters of Galaxies} \citep[see][]{zwicky_catalogue_1968}. This galaxy cross-IDs include SDSS J094401.86-003832.1\footnote{The galaxy SDSS Plate-MJD-Fiber reference is 266-51630-100} from the \emph{Space Digital Sky Survey} \citep[see][]{ahumada_16th_2020}, SHOC270 from the the \emph{SDSS HII-galaxies with Oxygen abundances Catalog} \citep[see][]{kniazev_strong_2004} and MCG +00-25-010 from the \emph{Millennium Galaxy Catalogue} \citep[see][]{propris_millennium_2007}. The analysis is based on the Integrated Field Unit data cube from the MUSE (Multi Unit Spectroscopic Explorer) instrument \citep[see][]{bacon_muse_2010}. In addition to the chemical analysis, we take this opportunity to present novel techniques for the treatment of Big Data observations. Finally, the authors discuss the discrepancies between the direct method and the photoionization model fitting. Both analysis make use of a neural networks sampler to fit the observed fluxes. The conclusions reached provide useful guidelines in how to properly consider the flux, and its uncertainty, on Bayesian photoionization model fitting. 

This manuscript has the following structure. Section \ref{sec:data_methods} presents the observational data and the tools employed to analyse its continuum and emission features. Section \ref{sec:analysis-and-results} continues with the results from the direct method and photoionization modelling along with a brief methodology description. The discussion starts at section \ref{sec:direct_discussion} interpreting CGCG007-025 physical properties. In section \ref{sec:model_fitting_disc}, we debate the impact on the methodology design on the photoionization modelling measurements. Finally, the main conclusions are summarized in section \ref{sec:conclusion}.

\pagebreak

\section{Observational data and methodology} \label{sec:data_methods}

\begin{figure}
\includegraphics[width=1.0\columnwidth]{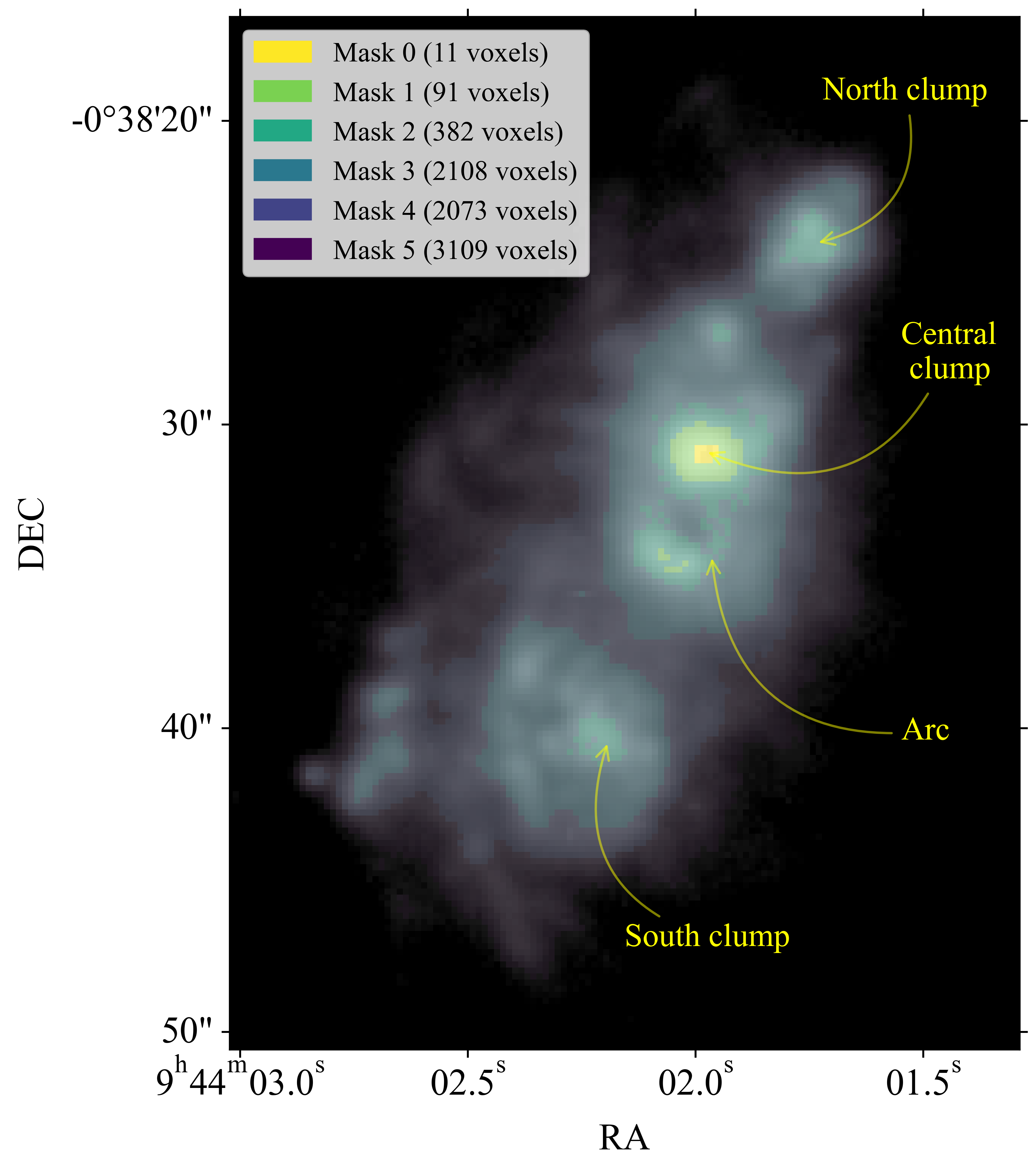}
\caption{\label{fig:HalphaMasks} Continuum-subtracted $H\alpha$ map from the MUSE observation of CGCG007-025. The over-plotted shaded regions correspond to the spatial masks described in section \ref{sec:data_methods}.}
\end{figure}

\begin{figure}
\centering
\includegraphics[angle=90,height=0.90\textheight]{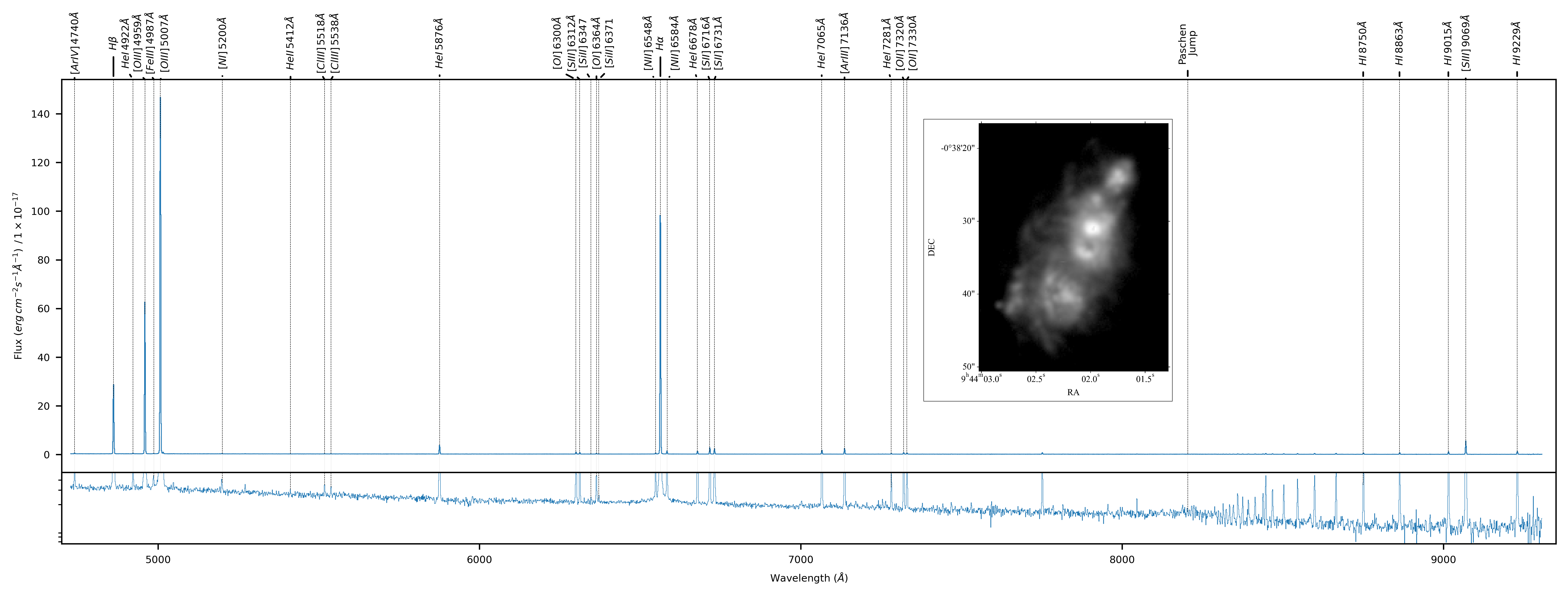}
\caption{\label{fig:voxel_spec} Spectrum from a voxel in the central core of CGCG007-025 with array coordinates 167-170. The lower panel is displayed in logarithmic scale with a constrained flux limits for a better focus on the continuum. The image corresponds to the $H\alpha$ band from the MUSE cube. The emission lines are labeled using the \textsc{lineidplot} library by \citet{nair_lineid_2016}}
\end{figure}

\begin{table}
\caption{\label{tab:emission_fluxes_voxel_core}Emission line fluxes for the voxel 167-170 at the central CGCG007-025 cluster. Column (1) Wavelengths in $\AA$ at rest frame. The labels with a "g" subscript identify emission lines whose measurement consisted in a Gaussian profile deblending. The "w1", "w2" suffixes refer to additional kinematic components on the transitions.  Column (2) Reddening curve. Columns (3-4) display the observed $F(\lambda)$ and extinction corrected $I(\lambda)$ fluxes relative to $F(H\beta) = I(H\beta) = 1000$. At the bottom of the table the logarithmic extinction coefficient, the equivalent width and the observed flux for $H\beta$ are tabulated (the flux units are $10^{-17} erg\,cm^{-2} s^{-1}$).}
\centering{\input{tables/example_emission_lines}}
\end{table} 

The current analysis on CGCG007-025 is based on observations made with the ESO (European Southern Observatory) VLT (Very Large Telescope) facility at the La Silla Paranal Observatory under programme ID 0102.B-0325. The data cube was obtained the $27^{th}$ of February 2019. The MUSE instrument wide field mode set up covered a $1 \,arcmin^{2}$ field in a $316\times328$ voxel (volumetrix pixel) array, where the $H\alpha$ emission extends over 7700 voxels. The SDSS observation of CGCC007-025 predicts a galactic redshift of $z=0.00469$ resulting in a distance of 26.7 Mpc. This translates into a voxel spatial resolution around the 100 parsecs.

The data was downloaded from the ESO archive already reduced with the MUSE pipeline \citep[][, version 1.2]{weilbacher_design_2012}. The wavelength range is $4750-9349\AA$ with a constant spectral resolution of $1.25\AA$ per pixel $\left(R_{H\alpha}=5250\right)$. The fits file is opened with the instrument \textsc{MPDAF} (Muse Python Data Analysis Framework) library by \cite{bacon_mpdaf_2016}.

Several pixels in the $H\alpha$ band had non-a-numerical (NaN) value in the 5 most intense voxels of the central clump. Attempting to interpolate the missing entries resulted in relative $H\alpha$ fluxes, below the expected emissivity $H\alpha/H\beta=2.86$ ratio. Moreover, in the central core of CGCG007-025 the $[OIII]5007\AA/[OIII]4959\AA$ flux ratio is below the theoretical emissivity ratio $(\approx 2.97)$. This photons loss could be explained by the saturation of the CCD (Charge Coupled Device) or a loss in the CCD linearity due to the intense radiation from these transitions.

To deal with these issues, the cube voxels were labelled with a set of masks to tailor the chemical analysis for the available lines. These masks are over-plotted on the CGCG007-025 $H_{\alpha}$ image in Fig.\ref{fig:HalphaMasks}. Regions 0, 1 and 2 correspond to the 99.99, 99.90 and 99.50 flux percentiles of the $[SIII]6312\AA$ band, our temperature diagnostic transition. Region 0 has the 11 most intense emission voxels, including those with the $H\alpha$ NaN pixels. Region 1 covers 91 voxels, where the $O^{2+}$ lines flux ratio is below the expected theoretical value. Finally, region 2 covers the remaining 382 voxels where the $[SIII]6312\AA$ has a good S/N ratio for a direct method analysis. Region 4, 5 and 6 correspond to the 97.50, 95.50 92.50 flux percentiles of the $H\alpha$ band. These regions have 2108, 2073 and 3109 voxels respectively.

\subsection{Emission line measurement}\label{sec:emission_fitting}

The emission flux measurements and the Gaussian profile fittings were computed with \textsc{LiMe}, the LIne MEasuring library (Fern\'andez et al, in prep). This python-based package has been developed by the authors for the analysis of large spectra samples with complex line profiles. Since this work represents its beta release\footnote{The library documentation and resources can be found at \href{https://lime-stable.readthedocs.io}{https://lime-stable.readthedocs.io}}, the following paragraphs describe its workflow:

\begin{enumerate}
  \item The user introduces a line spectral masks list. This is a 7 column table text file. The first column contains the line label (i.e. $H1\_6563A$). The remaining columns provide the wavelength values (in the rest frame), which are used to index the line location and two adjacent continua bands. These wavelengths are sorted from lower to higher wavelength values. This design adheres to the Lick indexes \citep[see][]{worthey_comprehensive_1994} for the analysis of stellar spectra. As we move to regions with lower ionization, the input masks exclude weaker emission lines in order to avoid false detections in noisier spectra.
  \item The galaxy continuum is fitted in an iterative process with a polynomial function. At each step, pixels deviating from the fit (including those in emission/absorption lines) are excluded to better characterise the object continuum. This calculation takes into account the pixel uncertainty if provided by the user. The output polynomial is used to normalise the input spectrum. This baseline is introduced in a peak detection algorithm \citep[see][]{yang_comparison_2009} from the \textsc{Scipy} library by \citet{virtanen_scipy_2020}. This provides the flux peaks and/or troughs pixels on the spectrum. In this work, the detection threshold was 3 times the normalized flux uncertainty for the emission lines and 2 for the absorptions lines. 
  \item The algorithm compares the input mask line regions with the peak/trough indexes found on the voxel spectrum (taking into consideration the galaxy redshift). In those cases with a match, a positive detection is assumed. For single lines, the algorithm adjusts the line mask to the FWZI (Full Width Zero Intensity) using the adjacent continua bands to compute the line continuum. 
  \item Among other parameters, each detected line has two flux calculations: The first one consists in one thousand loop Monte Carlo integration, while the second is a single/multi-Gaussian profile fitting. In both computations, the algorithm includes the sigma uncertainty spectrum from the current voxel to propagate the error into the measurements. For the chemical analysis, the integrated flux is used for single lines while for blended lines the Gaussian flux is used.
  \item The profile fitting is handled by the \textsc{LmFit} library by \citet{newville_lmfit_2014}. Our algorithm, however, does not use the default Gaussian parametrisation in \textsc{LmFit}. Instead, we define a model where $A_{i}$ is the Gaussian peak height (with respect to the line continuum), $\mu_{i}$ is the peak central wavelength and $\sigma_{i}$ is the Gaussian curve standard deviation. The subscript $i$ corresponds to the Gaussian component index. This flux is computed from the theoretical relation $\left(A_{i} \cdot \sqrt{2 \pi}\cdot\sigma_{i}\right)$.
  \item The library allows the user to define many constraints for the analysis of multiple kinematic components. However, in this observation of CGCC007-025 the only blended lines were a wide component in $H\beta$, $[OIII]4959\AA$, $[OIII]5007\AA$ and two in $H\alpha$, at the central clump. In the region 2 voxels, no wide components are observed and every line is assumed to have a single kinematic component.
\end{enumerate}
  
The \textsc{LiMe} script took 93 minutes to measure almost 61800 lines in the masked 7774 voxels (in a 3.6-GHz AMD-Ryzen7-3700X desktop). In the central clump of CGCC007-025, up to 45 lines are measured per voxel. This is the case of the voxel spectrum in Fig.\ref{fig:voxel_spec} corresponding to the central core of region 0. Table \ref{tab:emission_fluxes_voxel_core} shows the line fluxes and extinction corrected intensities for the same voxel. In contrast, most of the voxels in regions 3, 4 and 5 only display 4 lines.
The line measurements are stored as tables in a fits file, using the voxel coordinates (from the cube rectangular array) to label each extension. This file has a 18 MB compressed size and it is available on the manuscript online support material. The reader is encouraged to check the \textsc{LiMe} documenation at \href{https://lime-stable.readthedocs.io/}{lime-stable.readthedocs} regarding the physical/mathematical description of these measurements.

\subsection{Stellar population synthesis}
The stellar population synthesis on CGCG007-025 continuum was done with \textsc{FADO} by \citet{gomes_fitting_2017} and 
\textsc{Porto3D} by \citet{papaderos_nebular_2013,gomes_warm_2016} which invokes the population synthesis code \textsc{Starlight} by \citet{fernandes_semi-empirical_2005}.  The Simple Stellar Population (SSP) library uses the \citet{chabrier_galactic_2003} Initial Mass Function (IMF) with the Padova 1994 tracks \citep[see][]{gustavo_stellar_2001}. This treatment takes into consideration the nebular continuum in those voxels with a very young star forming region. This is computed from the intensity of the emission features. A very similar analysis was performed in \citet{fernandez_new_2021} for the analysis of long-slit spectra of Green Pea galaxies. This work compared the nebular continua computed by \textsc{FADO} with an iterative calculation of the nebular continuum taking consideration the gas chemistry. Both techniques produced consistent results. 
A future manuscript, (S\'anchez-Janssen et al, in preparation) will focus on the star formation history of CGCG007-025 and discuss the implications from these measurements. In the current work, however, these fittings are just used to measure the absorption from the underlying stellar population. This is done to better characterise the extinction from the hydrogen lines and the abundance from the helium transitions. The absorption lines measurements was done using \textsc{LiMe}. The workflow is very similar to the steps in section \ref{sec:emission_fitting}. 

\begin{table}
\caption{\label{tab:absorptions} CGCG007-025 mean absorption intensity relative to $H\beta$. Each transition includes the number of voxels where the absorption line was detected in the fittings of the stellar populations synthesis.}
\centering{\input{tables/absorptions}}
\end{table}

The $H\alpha$ and $H\beta$ absorption lines can be measured in the complete galaxy. However, the remaining HI and HeI lines are either too weak or non-detected in a large fraction of the voxels. Even in the low ionization regions, the emission component is at least one order of magnitude above the absorptions fluxes. The relative absorption of the recombination features, with respect to $H\beta$, shows a normal distribution for all the lines and all the spatial masks. Table \ref{tab:absorptions} shows the mean and standard deviation for the hydrogen and helium lines relative to the $H\beta$. The number of voxels used to calculate each value is displayed in the third column. These constrains lead to the following strategy in the extinction and chemical analysis: In the case of $H\alpha$ and $H\beta$, the emission fluxes are corrected by the absorption flux measured at each voxel. For the remaining lines, we use the averaged values from Table \ref{tab:absorptions}. In both scenarios, we propagate the error into the emission line flux. 

\section{Analysis and results} \label{sec:analysis-and-results}

\begin{figure}
\includegraphics[width=1.0\columnwidth]{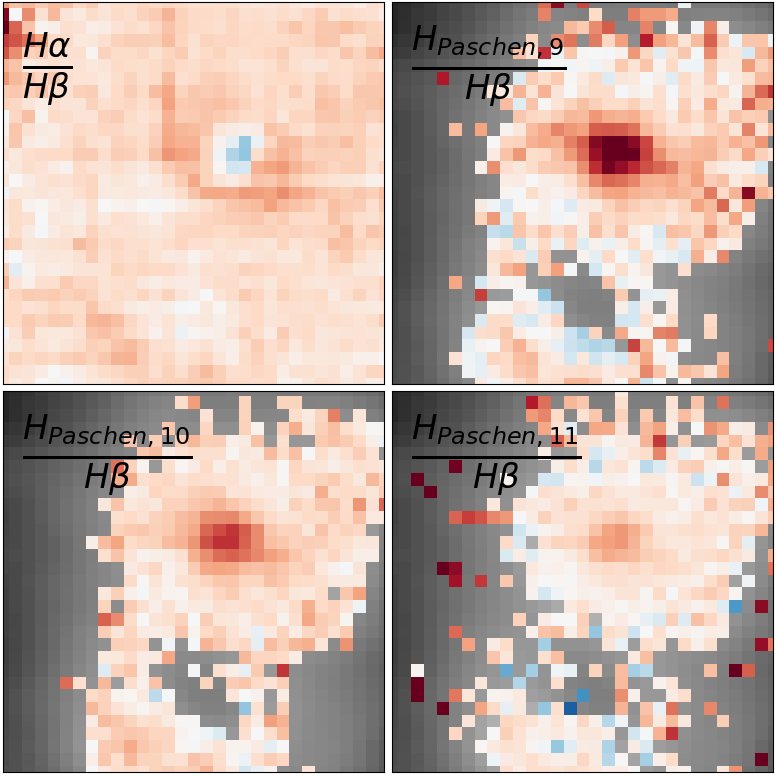}
\caption{\label{fig:HI_recombRatios}  Hydrogen line fluxes relative to $H\beta$ over the central cluster of CGCG007-025. The voxels with successful line detection are color coded: White voxels are close to the theoretical emissivity ratio, while red/blue voxels are above/below respectively. The drop in $H\alpha$ flux at the galaxy core is due to non-numeric pixels at the line. The background grey-scale voxels correspond to the $H\alpha$ band.}
\end{figure}

In the previous section, we described how there are drops on the $[OIII]5007\AA$ and $H\alpha$ fluxes in regions 0 and 1. For security's sake, both lines are excluded from the chemical analysis of the 102 voxels covered in these spatial masks. In the calculations, where the $[OIII]5007\AA$ flux is necessary, the value used is $2.97\cdot[OIII]4959\AA$. Similarly, instead of using the $H\alpha$ flux in the in the logarithmic extinction coefficient calculation, we use $H\beta$ and the brightest Paschen lines in our wavelength range: $HI_{P12\,}8750\AA$, $HI_{P11\,}8863\AA$, $HI_{P10\,}9015\AA$ and $HI_{P9\,}9229\AA$. The spatial plots of these lines fluxes (normalized by $H\beta$) shows a consisting pattern with higher ratios at the central core of CGCG007-025. This can be appreciated in Fig.\ref{fig:HI_recombRatios}. In region 2, both $H\alpha$ and $[OIII]5007\AA$ are included in the analysis while the Paschen lines are excluded due to their low signal-to-noise ratio. 

Table \ref{tab:atomic-data} at the appendix includes the references for the ionic species measured in this study. The transitions emissivity are calculated using \textsc{PyNeb} by \citet{luridiana_pyneb:_2015}. The following subsections provide a brief recap of the methodologies employed as well as the tabulated results. The data corresponding to individual voxels can be found on the online support material.

\subsection{Extinction} \label{sec:extinction}

\begin{figure}
\includegraphics[width=1.0\columnwidth]{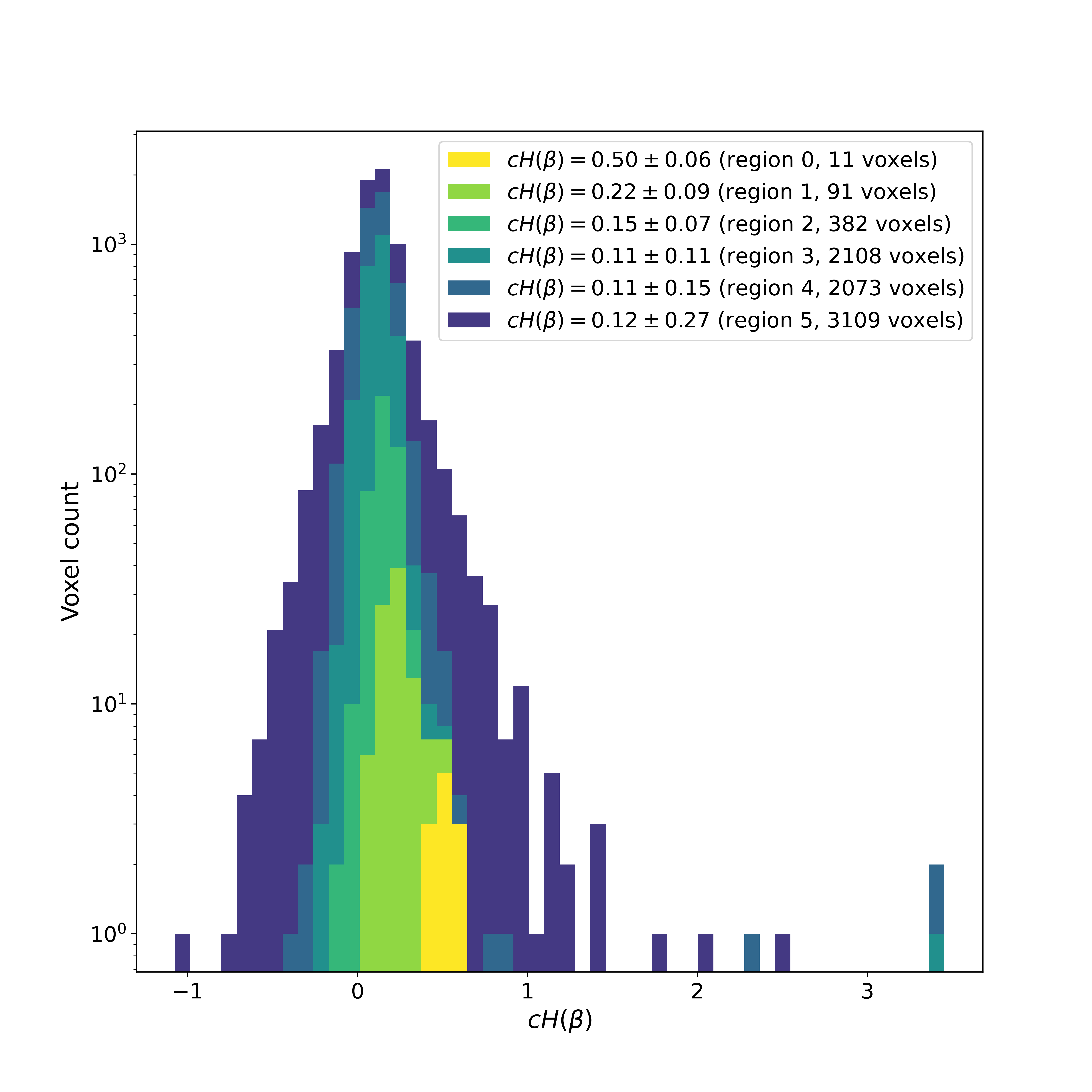}
\caption{\label{fig:extinc_distribution}Logarithmic stacked histogram with the voxel logarithmic extinction coefficient, $cH(\beta)$ spatial distribution. The histogram is color coded according to the spatial mask region. The legend includes the number of voxels at each spatial mask alongside the mean $cH(\beta)$ from the region voxels}
\end{figure}

The average Large Magellanic Cloud (LMC) reddening law by \citet{gordon_quantitative_2003} was selected for this study, assuming $R_{V}=3.1$. This curve was deemed appropriate for a very young star forming region. The hydrogen emissivites are calculated assuming a uniform spatial temperature and density of $10000\,K$ and $100\,cm^{-3}$ respectively. 

The mean voxel extinction of CGCG007-025 presents a low dust content with $cH(\beta) = {0.11}\pm{0.11}$. However, the core clump has a distinctive higher value. Fig.\ref{fig:extinc_distribution} displays the $cH(\beta)$ histogram, color-coded according to the galaxy spatial masks. The brightest 11 voxels show a tight mean value of $cH(\beta)={0.50}\pm{0.06}$ which drops to $cH(\beta)={0.22}\pm{0.09}$ for the 91 voxels in region 1. 
In the outer regions of CGCG007-025, we start to see some negative values of $cH(\beta)$. This can be explained by the low degree of ionisation: As $H\alpha$ and $H\beta$ become weaker, the observational noise affects the quality of the line flux measurement. Moreover, in the outskirts, the strength of the underlying stellar population absorption starts to be of the same order as the gas emission. This makes the accuracy of this correction more critical. These two parameters contribute to a poor linear regression on the $cH(\beta)$ calculation. Despite this, even in region 5, the extinction coefficient distribution is still centered at $cH(\beta)={0.12}\pm{0.27}$.

\subsection{Direct method physical properties}

The direct method is an observation-based treatment, where the ionic abundances are calculated directly from the observed fluxes. This approach assumes some simplifications on the ionization structure to characterise the electron temperature $(T_{e})$ and density $(n_{e})$. These parameters can be measured from the photons flux ratio of transitions, which are sensitive to them. However, these auroral lines are very weak. Consequently, this methodology usually imposes a hard constraint on the object selection due to the signal-to-noise requirement. In the case of CGCG007-025, the direct method is limited to 484 pixels from regions 0, 1 and 2.

In \citet{fernandez_bayesian_2019}, the authors presented a novel approach to fit the complete direct method parameter space simultaneously. This model consisted in one electron density, two electron temperatures, the logarithmic extinction coefficient, the optical depth for the HeI transitions and nine ionic species. An important advantage of this methodology was a more robust error propagation for the parameters, as well as, a faster treatment. The following paragraphs provide a brief summary: 

\begin{itemize}
    \item The ionic transitions emissivities are interpolated at  every step as a function of the sampled temperature and density by the \textit{RegularGridInterpolator} function from the \textsc{exoplanet} library by \cite{foreman-mackey_exoplanet_2021}.
    
    \item The solution sampling follows a Bayesian paradigm, where the prior distributions can be found in Table \ref{tab:Priors-and-likelihood}. The posterior solution is explored by a No-U-Turns (NUTs) sampler from the  \textsc{PyMC3} library by \citet{salvatier_probabilistic_2016}. The likelihood is defined with a Gaussian distribution for each line. These distributions standard deviation is the uncertainty of the emission line fluxes. Consequently, lines with higher error result in wider likelihoods, assigning flatter probabilities to the parameters involved on line flux. The likelihoods center is computed from the theoretical fluxes at the sampled coordinate using the relation:
    \begin{equation}
    \frac{F_{X^{i+},\,\lambda}}{F_{H\beta}}=X^{i+}\frac{\epsilon_{X^{i+},\,\lambda}\left(T_{e},\,n_{e}\right)}{\epsilon_{H\beta}\left(T_{e},\,n_{e}\right)}\cdot10^{-c\left(H\beta\right)\cdot f_{\lambda}}\label{eq:fluxFormula}
    \end{equation}
    where $\nicefrac{\epsilon_{X^{i+},\,\lambda}}{\epsilon_{H\beta}}$ is the relative emissivity at the transition wavelength $\lambda$, for an ion with abundance $X^{i+}$, for an electron  density $n_{e}$ and temperature $T_{e}\,\left(K\right)$. $c\left(H\beta\right)$ is the logarithmic extinction coefficient at $H\beta$ and $f_{\lambda}$ is the reddening curve.
    
    \item A common electron density is assumed for all the species. This parameter is anchored by the $S^{+}$ emission from the $[SII]6716,6731\AA$ doublet. 
    
    \item The chemical model considers two electron temperatures for a low and high ionization region, $T_{low}$ and $T_{high}$ respectively. In the high ionization region the $Ar^{3+}$, $O^{2+}$ and $y^{+}$ ions are located, while the remaining species belong to the low ionization region. The low ionization temperature is derived from the from the $[SIII]6312\AA$ auroral line. Even though, the $S^{2+}$ belongs to a midway ionization regime, the fast transition in the $S^{+}$ and $S^{2+}$ fractions, makes this a reasonable assumption \citep[see ][Fig.4]{fernandez_determination_2018} In our observation wavelength range, there are not suitable diagnostics for the high ionisation temperature. As a proxy, we are using the empirical relation from \citet{hagele_temperature_2006} to determine the $O^{2+}$ temperature:
    \begin{equation}
    T_{high}=T_{[OIII]}=0.8403\cdot T_{[SIII]}+2689\thinspace\left(K\right)
    \label{eq:TSIII-TOIII-relation}
    \end{equation} 
\end{itemize}

The final model consists in 11 dimensions: $n_e$, $T_{low}$, $c(H\beta)$ and eight ionic species ($Ar^{2+}$, $Ar^{3+}$, $N^+$, $O^+$, $O^{2+}$, $S^+$, $S^{2+}$ and $y^+$). It needs to be emphasized that even though our model considers two ionic temperatures, $T_{high}$ is not a free parameter. Indeed, due to the available lines, only $T_{low}$ can be measured from the observed data, while $T_{high}$ is computed from eq.\ref{eq:TSIII-TOIII-relation} at each sampling step. Consequently, the model outputs do not include a $T_{high}$ distribution. 

\subsubsection{Electron densities, temperatures and $c(H\beta)$ distributions}

\begin{table*}
\caption{\label{tab:ne_Te_cHbeta} CGCG007-025 chemical properties from the neural direct method fitting. The columns corresponds to results from all, region 0, region 1 and region 2 voxels respectively. The tabulated values represent the median measurements distribution and the $16^{th}$-$84^{th}$ percentiles. The values in brackets correspond to the mean parameter uncertainty in the measurements. The chemical model computes the high ionization temperature $\left(T_{high}=T_{[OIII]}\right)$ using eq.\ref{eq:TSIII-TOIII-relation}. The metal abundances in the $12+log\left(\nicefrac{X}{H}\right)$ scale.} 
\centering{\input{tables/direct_method_ionic_abundances}}
\end{table*}

\begin{table*}
\caption{\label{tab:methods_comparison} Chemical spatial distributions for the methodologies discussed in the text. The first row contiains the elemental abundances from the direct method. The following columns contain the measurements for the $12+log\left(O/H\right)$ abundance, the $N/O$ excess and the ionization parameter $log(U)$ for the Photoionization model fitting techniques. The columns corresponds to results from the total, region 0, region 1 and region 2 voxels respectively. The tabulated values represent the median measurements distribution and the $16^{th}$-$84^{th}$ percentiles.} 
\centering{\input{tables/methodology_results}}
\end{table*}

Table \ref{tab:ne_Te_cHbeta} displays the $c(H\beta)$ coefficient, the electron temperatures and densities from the direct method fitting using the neural networks sampler. The first column corresponds to the distribution from all the voxels, while the subsequent columns are limited to the spatial masks displayed in Fig \ref{fig:HalphaMasks} and described in section \ref{sec:data_methods}. 

The logarithmic extinction coefficients seem to be slightly higher than those from those linear regression on hydrogen recombination coefficients. However, these small discrepancies are within the measurement uncertainty for all the regions considered. The deviation can be explained by the different model: In the extinction linear regression the emissivites are calculated using constant $T_e = 10000\,K$ and $n_e=100\,cm^{-3}$ values. In contrast, the direct method fits all the parameters simultaneously, included the temperature and density. In \citet{fernandez_bayesian_2019}, we argued that the slightly higher uncertainty for extinction coefficients from this approach could be due to an uneven spatial dust  distribution: While the hydrogen lines cover a very wide volume, the metals emission is expected to be closer to the new born stars. If the latter region has different dust properties, including the metals emission could bring a bias. However, as we can see from this spatial study, the $c(H\beta)$ values from both techniques have consistent nominal values and uncertainties. 

The low ionization temperature of CGCG007-025 is $T_{low} = 14776\pm1336\,K$. This temperature distribution profile is Gaussian-like for all the regions and the increase in uncertainty as we move towards the outskirts of CGCG007-025 clusters can be explained by the larger uncertainty in the $[SIII]6312\AA$ flux measurement. The high ionization temperature for CGCG007-025 derived from eq.\ref{eq:TSIII-TOIII-relation} is $T_{high} = 15105\pm1122\,K$. At this regime, there is a high linearity between the $O^{2+}$ and ${S^{2+}}$ temperatures \citep[see][]{garnett_electron_1992,  hagele_temperature_2006,perez-montero_deriving_2014, fernandez_determination_2018}. Our model, however, does not take into consideration the uncertainty on the linear relation coefficients. 

Overall, the electron density derived from the $S^{+}$ doublet remains uniform across the brightest clusters of CGCG007-025. The mean density from the 484 voxels is $n_e = 107^{84}_{47}\,cm^{-3}$. This is the canonical value for the gas in low metallicity star forming regions as there is a well, on the photoionization equilibrium \citep[see][]{osterbrock_astrophysics_1974}. However, there is a significant density rise at the central, brightest cluster of the galaxy with a $n_e = 378^{34}_{63}\,cm^{-3}$ distribution. This density decreases to $n_e = 177^{84}_{64}\,cm^{-3}$ for region 1.

\subsubsection{Ionic and total abundances}

The ionic abundance distributions are tabulated in Table \ref{tab:ne_Te_cHbeta} for the different regions considered. The higher abundance uncertainty in region 2 can be explained by larger error on the temperature determination as the auroral line has a lower signal-to-noise at the clusters edge. This can be appreciated from the values in brackets which correspond to the mean parameter uncertainty for the considered voxels. Thus the width on the ionic abundances distribution can be partially explained by the bias on the measurement precision rather than the existence of several chemical components.

In the case of oxygen, the total abundance can be computed directly from the $O^{+}$ and $O^{2+}$ abundances as these ions are the main components \citep[see][]{pagel_survey_1978}. For the sulfur abundance, the ionization potential of $S^{3+}$ is sufficiently low for a non negligible component at the hottest gas of the galaxy \citep[see][]{oey_calibration_2000}. To account for this contribution, we use the ionization correction factor, $ICF(S^{3+})$ presented in \citet{fernandez_primordial_2018}:
\begin{equation}
log\left(\frac{Ar^{2+}}{Ar^{3+}}\right)=a\cdot log\left(\frac{S^{2+}}{S^{3+}}\right)+b\label{eq:ICF_S3}
\end{equation}
where $a=1.162\pm0.006$ and $b=0.05\pm0.01$. In those voxels, where the weak $[ArIV]4740\AA$ line is not detected, it is assumed that the $S^{3+}$ fraction is negligible. 

For the nitrogen abundance calculation, we take the approximation that the $N^+/O^+$ ratio equals the $N/O$ ratio in the low metallicity regime \citep[see][]{garnett_nitrogen_1990}. Consequently, the nitrogen abundance is the result of multiplying this ratio by the oxygen abundance.

Due to the high ionization potential of $He^{+}$, the $y^{2+}$ abundance is very small and dominated by phenomena beyond the photoionization of stellar atmospheres. In \citet{fernandez_determination_2018} low metallicity HII galaxy sample, the $y^{2+}/y^{+}$ ratio was always below $3\%$. In our wavelength range, we do not have access to the $HeII4865\AA$ transition. However, at central region of CGCG007-025, it may be possible to distinguish the $HeII5412\AA$ emission from the 4-7 Pickering series. Computing the abundance from its flux (Table \ref{tab:emission_fluxes_voxel_core}) using $T_{high}$, the relative ionic fraction is $\nicefrac{y^{2+}}{y^{+}}\lesssim0.99$. A similar regime has been reported in the primordial helium galaxy sample by \cite{peimbert_primordial_2016}. Regarding, the neutral helium fraction, \cite{pagel_primordial_1992} argued that at the intense radiation fields encountered in low metallicity BCD, the neutral fraction is very small. The models presented by the authors stated that the $ICF(He)\approx1$ for $\eta < 0.9$.  As it will be discussed in \ref{sec:model_fitting_disc} this is the case of CGCG007-025. Consequently, it is reasonable to assume that $y \approx y^{+}$ .

Finally, the argon abundance can be characterised by the sum of the $Ar^{2+}$ and $Ar^{3+}$ fractions \citep[see][]{perez-montero_ionized_2017}. However, the weak $[ArIV]4740\AA$ line at the edge of our wavelength range has a weak signal-to-noise in our observations, as inferred from the skewed distribution on its abundance determination.

\begin{figure*}
\includegraphics[width=1.0\textwidth]{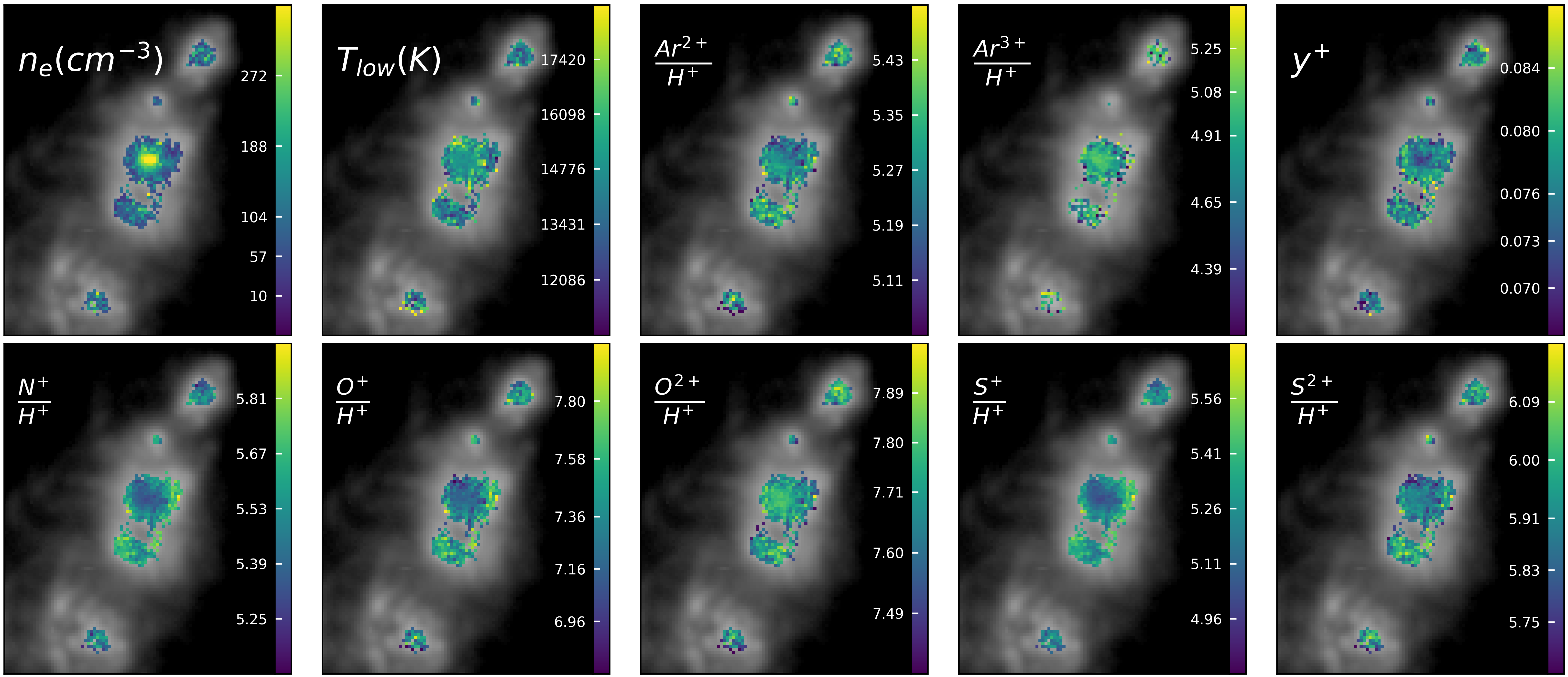}
\caption{\label{fig:param_maps} Voxel measurements for the $n_{e}$, $T_{low}$ and ionic abundances (in $12+log(X^{i+}/H)$ scale) from the direct method. The tabulated values on the color bars correspond to the median, $P_{16th}$-$P_{84^{th}}$ and $2 \times P_{16th}$-$P_{84^{th}}$. The $c(H\beta)$ maps are excluded from this figure since is distributions can be better appreciated from Fig.\ref{fig:extinc_distribution}. The background gray-scale voxels correspond to the MUSE $H\alpha$ band.}
\end{figure*}

The total abundances are tabulated in Table \ref{tab:methods_comparison} and Fig.\ref{fig:param_maps} displays the same results graphically over the CGCG007-025 $H\alpha$ map.

\subsection{Photoionization model fitting} \label{sec:photo_modellin}

Simulating the observable parameters from a grid of conditions, for both the ionizing source and the surrounding gas, provides astronomers with a powerfull tool to understand the star formation phenomena. In this work, we compare two techniques to fit models from the \textsc{PopStar} evolutionary synthesis grids by \citet{molla_popstar_2009, martin-manjon_popstar_2010}. These models were compiled by \citet{perez-montero_deriving_2014} with an initial mass function of \citet{chabrier_galactic_2003} and a burst age of $1\,Myr$. The model metallicity is scaled to the solar value with the oxygen abundance. The models parameterise the variation of the $N/O$ ratio from its dependence on the $[NII]$ line fluxes as the $N/H$ abundance changes. The electron density is constant with $100\,cm^{-3}$, as well as, the cluster age of $1\,Myr$. The grids were generated using \textsc{Cloudy} v17 by \citet{ferland_2017_2017} with the following ranges: $7.1 \leq 12+log(OH) \leq 9.0$, $-2.0 \leq log(N/O) \leq 0$ and $-4.0 \leq log(U) \leq -1.5$. As in the direct method treatment, we are excluding those lines suspected of saturation (see section \ref{sec:data_methods}).

\subsubsection{\textsc{HII-CHI-mistry} sampling} \label{sec:HII-CHI-mistry modelling}

The first tool to fit the photoionization model is \textsc{HII-CHI-mistry} v5.22. The complete algorithm description can be found in \citet{perez-montero_deriving_2014,perez-montero_bayesian-like_2019,perez-montero_extreme_2021}, but at its core, the sampling consists in three interpolations. The grid described in the previous subsection is treated as a 3-dimensional array. The first interpolation is across the $N/O$ axis as a function of the $[NII]6584\AA$ fluxes. The second interpolation anchors the $12+ log(O/H)$ abundance. Finally, the algorithm fits the ionization parameter, $log(U)$, according to the input lines.  Additionally, the algorithm includes several empirical models to constrain the sampled parameters, in those cases where not all the model input lines are available.

Table \ref{tab:methods_comparison} includes the results from the \textsc{HII-CHI-mistry} fittings. Each row corresponds to one of the model parameters, while the first column contains the average result from all the voxels while in the subsequent columns present the distributions from the spatial signal-to-noise masks described in \ref{sec:data_methods}. In addition to the \textsc{PopStar} configuration, the algorithm was configured to use the interpolation feature to increase the grid resolution. Moreover, the default Monte Carlo chain length was increased from 25 to 200. Unfortunately, due to the the partial overlap between MUSE wavelength range and \textsc{HII-CHI-mistry} input lines, only $[OIII]4959\AA$, $[OIII]5007\AA$, $[NII]6584\AA$, $[SII]6716\AA$ and $[SII]6731\AA$ are used as inputs. The line fluxes are corrected from the dust extinction using the voxel $c(H\beta)$ calculated in section \ref{sec:extinction}.

\subsubsection{Neural network sampling}\label{sec:neural_sampling}

The second approach to fit the photoionization model is using the neural networks sampler from the direct method treatment. This approach was first introduced in the chemical analysis of the three Green Peas galaxies in \citet{fernandez_new_2021}. At the time, it was argued that this scheme provided a faster treatment, while making possible the sampling of grids with a higher number of dimensions and input lines. At this point, we perform three quality checks on this methodology: 

\begin{figure*}
\includegraphics[width=1.0\columnwidth]{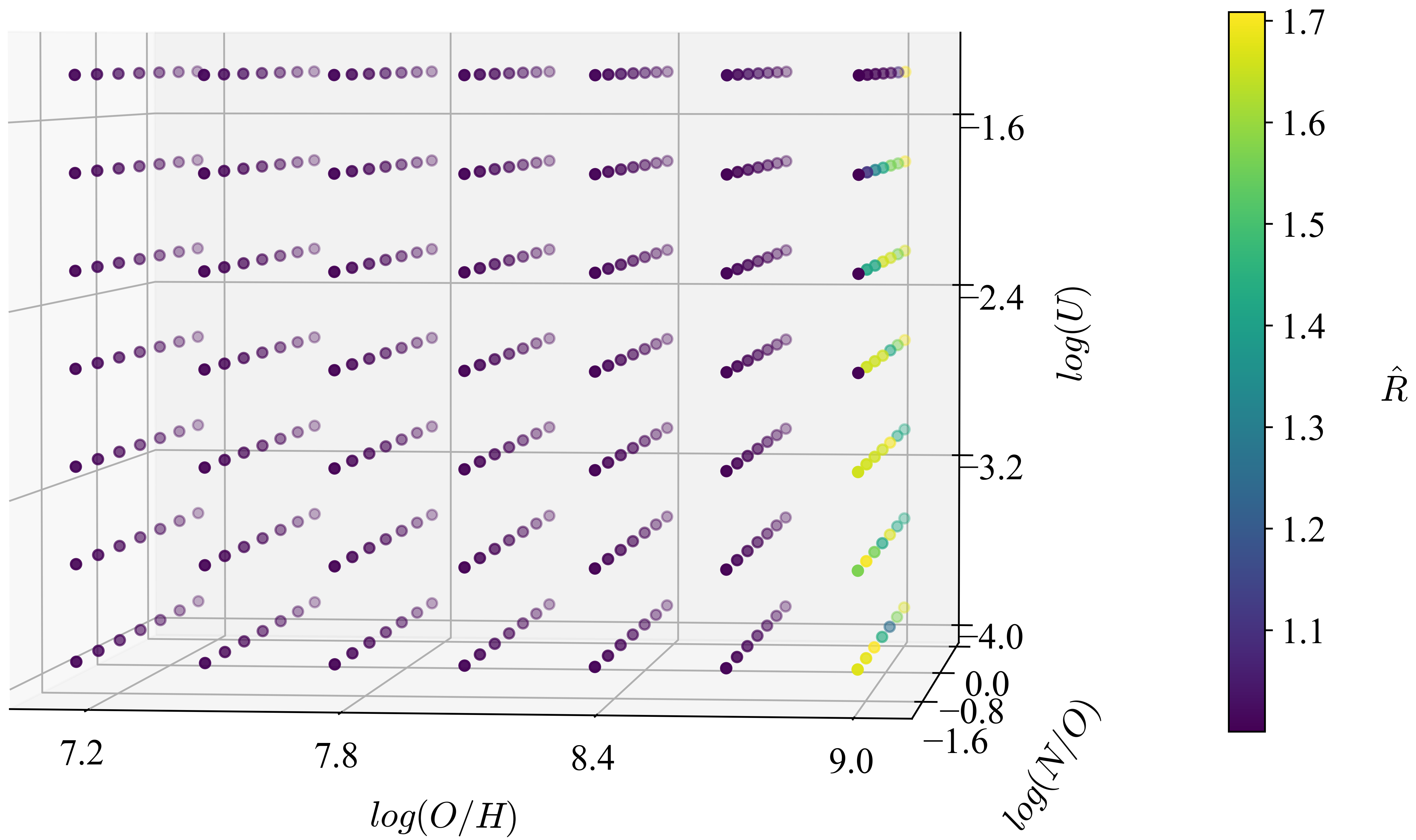}
\includegraphics[width=1.0\columnwidth]{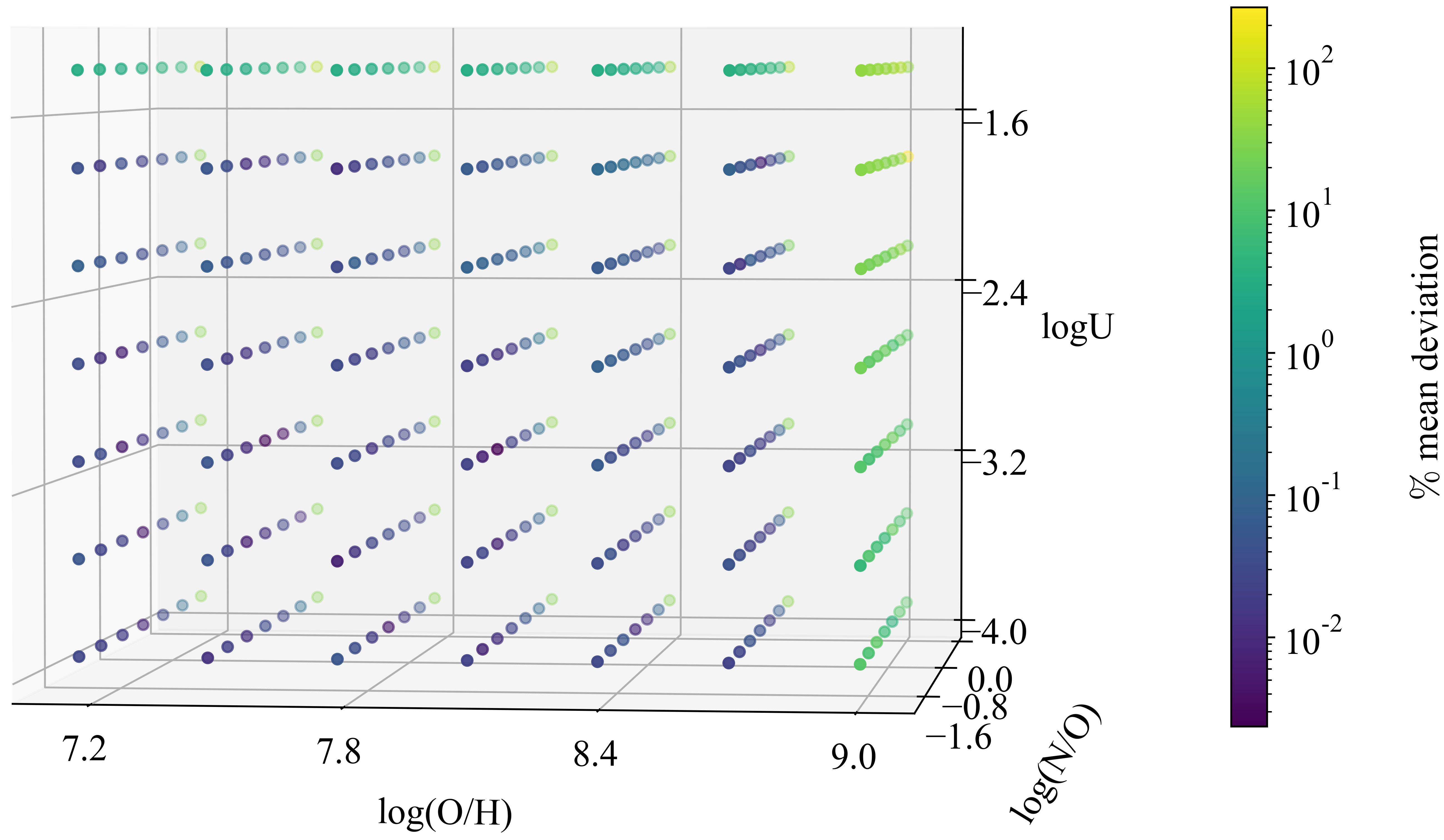}
\caption{\label{fig:scatter_3D} Scatter plots with the diagnostics for the neural photoionization fitting of the MUSE range lines from the \textsc{PopStar} grid described in section \ref{sec:photo_modellin}. Each point represents 10 fittings with a 2500 steps per thread. Left) The color scale represents the mean Gelman-Rubin $\left(\hat{R}\right)$ statistic calculated from the 3 parameter output chains. Fittings with with $\hat{R}>1.4$ have failed to converge. Right) The mean discrepancy for the three parameters with their true value.}
\end{figure*}

Firstly, we try to fit as many lines as possible from the complete photoionization grid of \textsc{HII-CHI-mistry} (\citep[][private communication]{perez-montero_extreme_2021}. Indeed, the number of input lines is almost the same as in the direct method treatment. Some of the lines not included are the hydrogen Paschen lines. However, since the fluxes are corrected from the dust extinction using the coefficients in section \ref{sec:extinction}, the hydrogen transitions do not supply additional information to the fitting. 

Secondly, we perform a systematic analysis on the algorithm convergence for the grid parameter space. At seven points per $12+log\left(O/H\right)$, $log(N/O)$ and $log(U)$ axis, we get the line fluxes within the MUSE wavelength range. Afterwards, these fluxes are introduced into the neural sampler to fit the model parameters. The results from these tests can be found in the 3-dimensional scatter plots in Fig. \ref{fig:scatter_3D}. Each point represents 10 fittings, one per processor-thread, consisting in 2500 steps Monte-Carlo chains. The first 500 steps are excluded from the measurement statistics, since they represent the time it takes the simulation to reach the solution coordinate. On the left hand side plot of Fig. \ref{fig:scatter_3D}, the dots color represent the Gelman-Rubin or $\hat{R}$ statistic \citep[see][]{gelman_inference_1992}. This is the ratio of the variance of the combined 10 chains (20000 points in this case) by the averaged variance of the individual chains. In an ideal output, this ratio should be one. In the \textsc{PyMC3} sampler, chains with $\hat{R}>1.4$ are labelled with an unsuccessful convergence. Similarly, the dot colors in the second graph quantify the 3 parameters average deviation from their true values. From these diagrams it can be concluded that the algorithm has in general, a very good convergence at the true values. The exception lies on the higher metallicity end. Looking at the individual traces \citep[check Fig. 4 in][]{fernandez_new_2021}, some of the chains are stuck grid edge values. Moreover, we find that while there is convergence at the lower $log(U)$ boundary, the code is not finding the true solutions. Consequently, parameter measurements from these regions in real spectra should be regarded with suspicion.

Finally, in this work we explore the impact of the input lines uncertainty on the photoionization model fitting. The tabulated data in Table \ref{tab:methods_comparison} show the results from different configurations. As in the direct method analysis, the Bayesian likelihood is defined with a normal distribution. This is further discussed in the next section. 

Unlike in the direct method analysis, the sampling uses an automatic differentiation variational inference (ADVI) to set the threads initial values. Testing the inbuilt algorithms in \textsc{PyMC}, this is the one with the best convergence for the parameter space.

\section{Discussion} \label{sec:direct_discussion}

The chemical analysis of CGCG007-025 shows a normal spatial distribution for most parameters. The exceptions to this pattern are the dust extinction and the electron density. The results are tabulated in Table \ref{tab:ne_Te_cHbeta} and Table \ref{tab:methods_comparison}. The same results are displayed graphically in Fig. \ref{fig:param_maps} and Fig. \ref{fig:maps_methodologies}. In the spectrophotometric catalogue of low metallicity compact HII galaxies of \citet{terlevich_spectrophotometric_1991}, the $c(H\beta)$ distribution of the 425 galaxies peaked in the $c(H\beta)\approx0.2-0.3$ range. This is the regime observed for most of the central clump (region 1) and north and south clumps (region 2). However, at the core of the central clump, where the brightest voxels are found, the extinction of the galaxy is almost three times higher $\left(c(H\beta)=0.52\pm0.06\right)$ that the mean value of the 73 voxels, $\left(c(H\beta)=0.21\pm0.08\right)$. Despite the dead and saturated-risk pixels at the $H\alpha$ line. This steep extinction gradient is observed in the 4 most intense Paschen lines ($HI_{P12\,}8750\AA$, $HI_{P11\,}8863\AA$, $HI_{P10\,}9015\AA$ and $HI_{P9\,}9229\AA$). This can be appreciated in Fig.\ref{fig:extinc_distribution}. Additionally, the same behaviour is observed in the electron density measurements. The average density of CGCG007-025, $n_{e} = 104\pm84$ $cm^{-3}$, is a standard value for low metallicity star forming galaxies \citep[see][]{hagele_precision_2008}. In contrast, the central knot has densities thrice times higher, $n_{e} = 378\pm63\ cm^{-3}$. 

Published studies on CGCG007-025 have found similar parameter values. While \citet{van_zee_evolutionary_2000} presented a photometric dwarf sample with both CGCG007-025 and its neighbour UGC5202, \citet{van_zee_oxygen_2005} focused on the gas compositions of dwarfs with intense star formation. The latter work included three slits on the central cluster of CGCG007-025. These extinctions were $c(H\beta)_{WN}=0.22\pm0.05$, $c(H\beta)_{W}=0.31\pm0.04$ and $c(H\beta)_{S}=0.12\pm0.05$. The authors manuscript did not tabulate the electron densities but from the reported [SII] ratios we can measure $n_{e, WN}=119\pm94$, $n_{e, W}=189\pm124$ and $n_{e, S}=162\pm118$ $cm^{-3}$. \citet{senchyna_ultraviolet_2017} studied the ultraviolet emission of CGCG007-025 (labelled as SBID 2) using the HST/COS observations. Their work also includes a chemical analysis with SDSS and MMT spectra. The authors presented an extinction coefficient of $c(H\beta)=0.21\pm0.07$ and an electron density of $n_{e}=186^{33}_{32}$ $cm^{-3}$. Izotov and collaborators have included CGCG007-025 in several of their papers focused on the primordial helium abundance determination. Finally, CGCG007-025 gas composition has been analysed by the CLASSY team.
\citet{arellano-cordova_classy_2022} explores the aperture effects on this sample by comparing twelve APO/SDSS long-slit spectra, seven LBT/MODS long-slit spectra and three LBT/MUSE IFU cubes and six Keck/ESI echelle spectra. The authors find that the density and extinction remain roughly uniform across the three data sets. Their IFU sub-sample does not include CGCG007-025. However, they warned that in their 3 IFUs, the extinction magnitude decreases up to 53\% from the center of the galaxy. Moreover, similar behaviour was observed while investigating the aperture extraction on the IFU $[SII]6716\AA,6731\AA$ doublet band. Higher values were found at the center of the galaxies. The extinction reported for CGCG007-025 was $c(H\beta)\approx0.24$ from the three longslit data sets. The reported mean density from the $[SII]$ lines was $\approx120\,cm^{-3}$. Additionally, the authors report the density from the $[OII]3727\AA,3729\AA$ doublet from extinction measured with the LBT/MODS spectrum. This value was considerably higher with $ne[OII]=480\pm170\,cm^{-3}$. This value was above the $100\,cm^{-3}$ mean discrepancy found between the $[OII]$ and $[SII]$ densities (the latter always lower) for the CLASSY sample. In conclusion, the literature results agree with the spatial distribution measured in this paper.

The cause behind the spatial variation on these parameters may be found in the evolutionary state of the central clump in CGCG007-025. The value of $c(H\beta)$ (as well as the color excess, $E_{B-V}$) depends on the grain properties (composition and size) and the column density in the line of sight \citep[see chapter 7 of][]{osterbrock_astrophysics_1974}. Assuming the same dust type, the higher extinction can be explained by a denser medium. In our Galaxy, the dust extinction per unit column density of hydrogen, $A_{V}/N(H)$, displays similar values from stellar samples as confirmed by \citet{bohlin_survey_1978} and \citet{rachford_molecular_2008}. As a loose proxy, we can use the $c(H\beta)/n_{SII}$ resulting in $0.0014$, $0.0015$ and $0.002$ for regions 0, 1 and 2. The spatial distribution of this ratio is rather uniform across the galaxy central clump. More-alike the behaviour of the other chemical measurements. CGCG007-025 was one of the test cases for the spectral synthesis library \textsc{FADO} by \cite{gomes_fitting_2017}. The fiber from the SDSS spectrum was centered at the central clump of CGCG007-O25 and the light-weighted age resulted in a value of $log(Age)_{L}=6.34\,yr$. Our stellar synthesis on the complete MUSE region agrees with this result, while confirming an older stellar population in the outer regions. Consequently, the density inhomogeneities can be explained by a very young cluster at the core. These measurements, however, will have their dedicated analysis in future work by (S\'anchez-Janssen et al, in preparation).

The previous bibliography includes the measurements of the electron temperature, mainly, from the auroral $[OIII]4363\AA$ line. In \citet{van_zee_oxygen_2005} three WN, W and S slits, the measurements were: $T[OIII]=14770^{530}_{500}\,K$, $16010^{340}_{330}\,K$, $13920^{500}_{430}\,K$. These results lead to an oxygen abundance of $7.83\pm0.03$, $7.78\pm0.03$ and $7.85\pm0.03$ respectively. In contrast, the analysis of \citet{senchyna_ultraviolet_2017}, combining the SDSS and the MMT fluxes (for the $[OII]3727,3729\AA$ doublet) resulted in $T[OIII]=15800\pm500\,K$ and $12+\log(O/H)=7.81\pm0.07$. The two long-slit spectra in \citet{izotov_systematic_2004} include both the $O^{2+}$ and $S^{2+}$ temperatures, where: $T[OIII]_{1}=16470\pm170\,K$, $T[OIII]_{2}=16560\pm260\,K$  and $T[SIII]_{1}=15370\pm140\,K$, $T[SIII]_{2}=15440\pm260\,K$. The reported oxygen and sulphur abundances were $12+\log(O/H)=7.775\pm0.010$, $7.738\pm0.013$ and $12+\log(S/H)=5.92\pm0.01$, $5.87\pm0.03$.  Finally, the mean CLASSY measurements from \citet{arellano-cordova_classy_2022} ground telescopes compilation have a mean temperature of $T[SIII]\approx14500\,K$ and $T[OIII]\approx15500\,K$. The mean oxygen abundance was $12+\log(O/H)\approx7.79$. The authors concluded that the aperture effects on the metallicity measurements is $<0.1\,dex$. Our results agree with this estimation.

The variability in the chemical content of CGCG007-025 found in the literature above \citep[and others not discussed, see][]{kniazev_strong_2004,guseva_balmer_2007,shirazi_strongly_2012,chevallard_physical_2018,kurichin_new_2021} can be explained by the spatial distribution observed in this work. In the case of the oxygen abundance, distribution measured are  $7.88\pm0.03$, $7.87\pm0.09$ are $7.88\pm0.12$ for the voxels in regions 0, 1 and 2 respectively. Within the core of the galaxy (the brightest knot with 11 voxels $\approx$ 48 pc radius), the ionic abundance the variation is below 0.05 dex. This sigma can be explained by the uncertainty on the flux measurement. Mainly, from the $[SIII]6312\AA$ temperature diagnostic line. However, once we start to consider the remaining 91 and 382 voxels (with radii $\approx$ 140 pc and 382 pc respectively), the dispersion on the parameters can no longer be explained by the flux measurement uncertainty. From the data in Table \ref{tab:ne_Te_cHbeta}, the 484 voxels ratio between their ionic abundances distributions sigma and their uncertainty distribution sigma is $\sigma_{measurement}/\sigma_{uncertainty}\approx2.1$. Looking at the spatial maps in Fig.\ref{fig:param_maps}, however, there isn't a distinctive gradient on these parameters (except for $c(H\beta)$ and $n_{e}$ as discussed previously). In well observed objects, such as our Galaxy or Messier33, a chemically homogeneous gas has been reported, \citep[see][respectively]{esteban_about_2022,rogers_chaos_2022}. This is because the galactic structure provides well-mixed enriched from inside out enrichment. In an compact system such as CGCG007-025, however, this points towards a very sudden starburst, fueled by fresh metal-poor gas.
 
This chemical distribution places CGCG007-025 within the Extreme Emission Line Galaxies (EELGs) family. These objects are characterised by their dominant nebular component (which hides the underlying galaxy stellar continuum), compact sizes and low metallicities \citep[see][]{smit_evidence_2014,amorin_evidence_2014,roberts-borsani_zgtrsim_2016,stark_ly_2017,calabro_characterization_2017,tang_mmtmmirs_2019}. These properties are expected to be more common in the early universe, \citep[see][]{atek_hubble_2014,amorin_analogues_2017,maseda_muse_2018,endsley_o_2021}. Nonetheless, in the nearby universe, $z\approx0.14-0.36$, Green Pea galaxies (GPs), are being targeted as analogues at the lower redshift \citep[see][]{cardamone_galaxy_2009}. These objects have small sizes, with few or below 1 kpc effective radius \citep[see][]{amorin_star_2012,yang_ly_2017} and stellar masses bellow $log(M_{*}) < 10M_{\odot}$. This is the case of CGCG007-025 whose reported neutral hydrogen mass is $log(M(HI))=8.62M_{\odot}$ \citep[see][]{van_zee_evolutionary_2001}, dynamical mass $log(M_{dyn}) = 9.36M_{\odot}$ \citep[see][]{van_zee_oxygen_2005} and stellar mass $log(M_{*}) = 8.17M_{\odot}$ \citep[see][]{gavilan_chemical_2013}. The oxygen abundance in these objects is around 20\% the solar value, $7.4<12+log(O/H)<8.4$ \citep[see][]{amorin_oxygen_2010,perez-montero_extreme_2021,fernandez_new_2021}. All these properties are within the gas emission characteristics measured here for CGCG007-025.

\subsection{Radiation field and photoionization model fitting implications} \label{sec:model_fitting_disc}

\begin{figure}
\includegraphics[width=1.0\columnwidth]{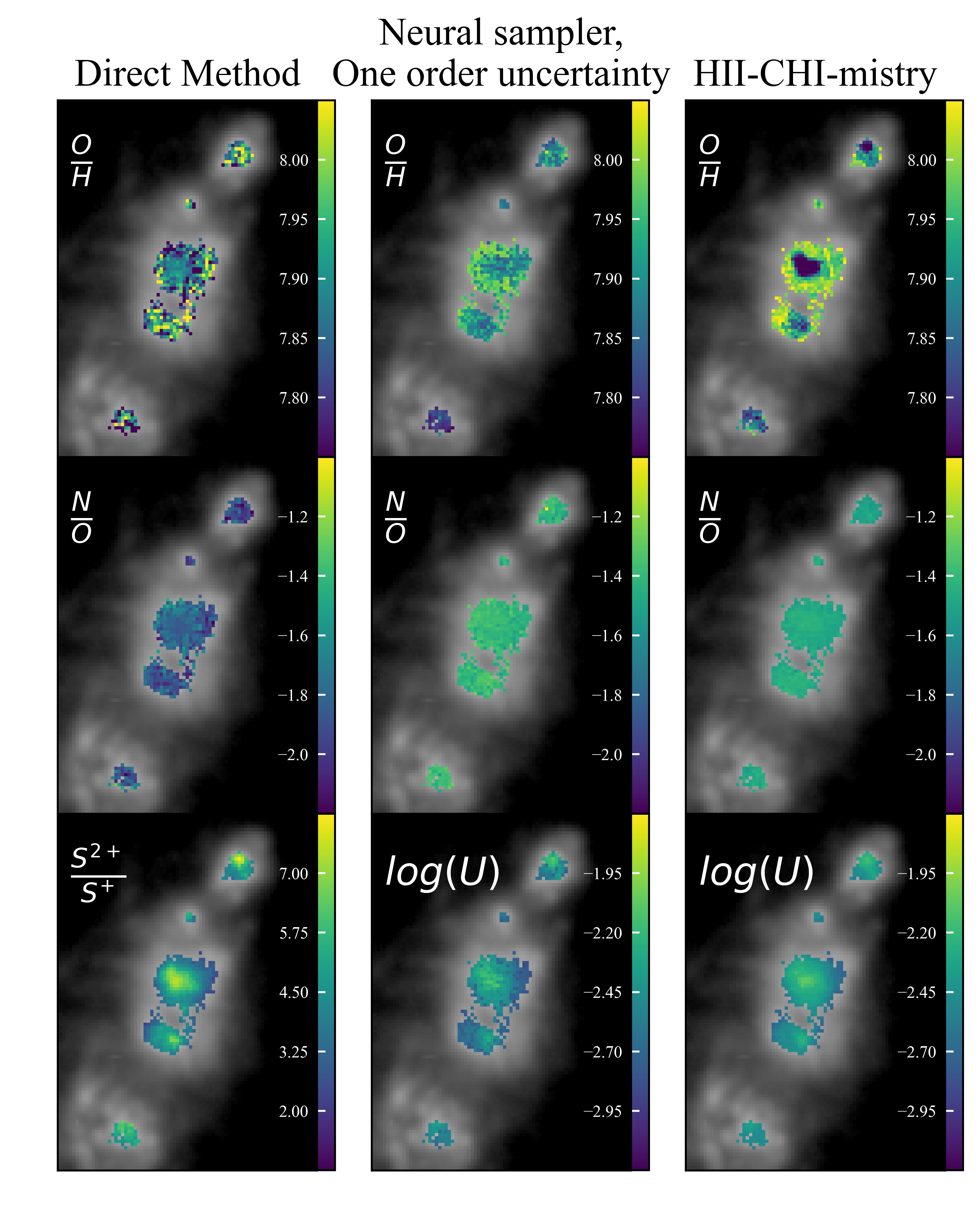}
\caption{\label{fig:maps_methodologies}Comparison between the photoionization model fitting techniques and the direct method. Instead of $log(U)$, we use $S^{2+}/S^{+}$ as a proxy in the direct method. In the oxygen distribution map, the color bar tabulated values are fixed. In the remaining maps, the color bars display the median, $P_{16th}$-$P_{84^{th}}$ and $2 \times P_{16th}$-$P_{84^{th}}$. The background gray-scale voxels correspond to the MUSE cube $H\alpha$ band.}
\end{figure}

In addition to the gas properties, the emission fluxes can provide constraints on the ionization source of a star forming region. This methodology is known as photoionization model fitting. \citet{stasinska_what_2007} describes it as the "royal" approach to analyse emission line spectra. Moreover, this review illustrates its common challenges: "How to explore the parameter space? How to deal with error bars? How to test the model validity?". While the latter issue is within the astrophysicist responsibilities, the former two are heavily affected by the researcher mathematical/computational resources. We take advantage of this work on CGCG007-025 to propose a to anchor the parameter space exploration and the error bar definition.

In section \ref{sec:neural_sampling}, we presented the sampling algorithm alongside the convergence tests for the observed transitions in the MUSE wavelength range. For these inputs, we tested the convergence stability for the models $log(U)$, $12+log(O/H)$ and $log(N/O)$ parameter space. At this point, knowing from the direct method the CGCG007-025 $12+log(O/H) = 7.88\pm0.10$ and $log(N/O)=-1.84\pm0.12$ distributions, we can expect the sampler to provide stable and accurate solutions for this galaxy (within the theoretical model assumptions). 

\begin{figure*}
\includegraphics[width=1.0\columnwidth]{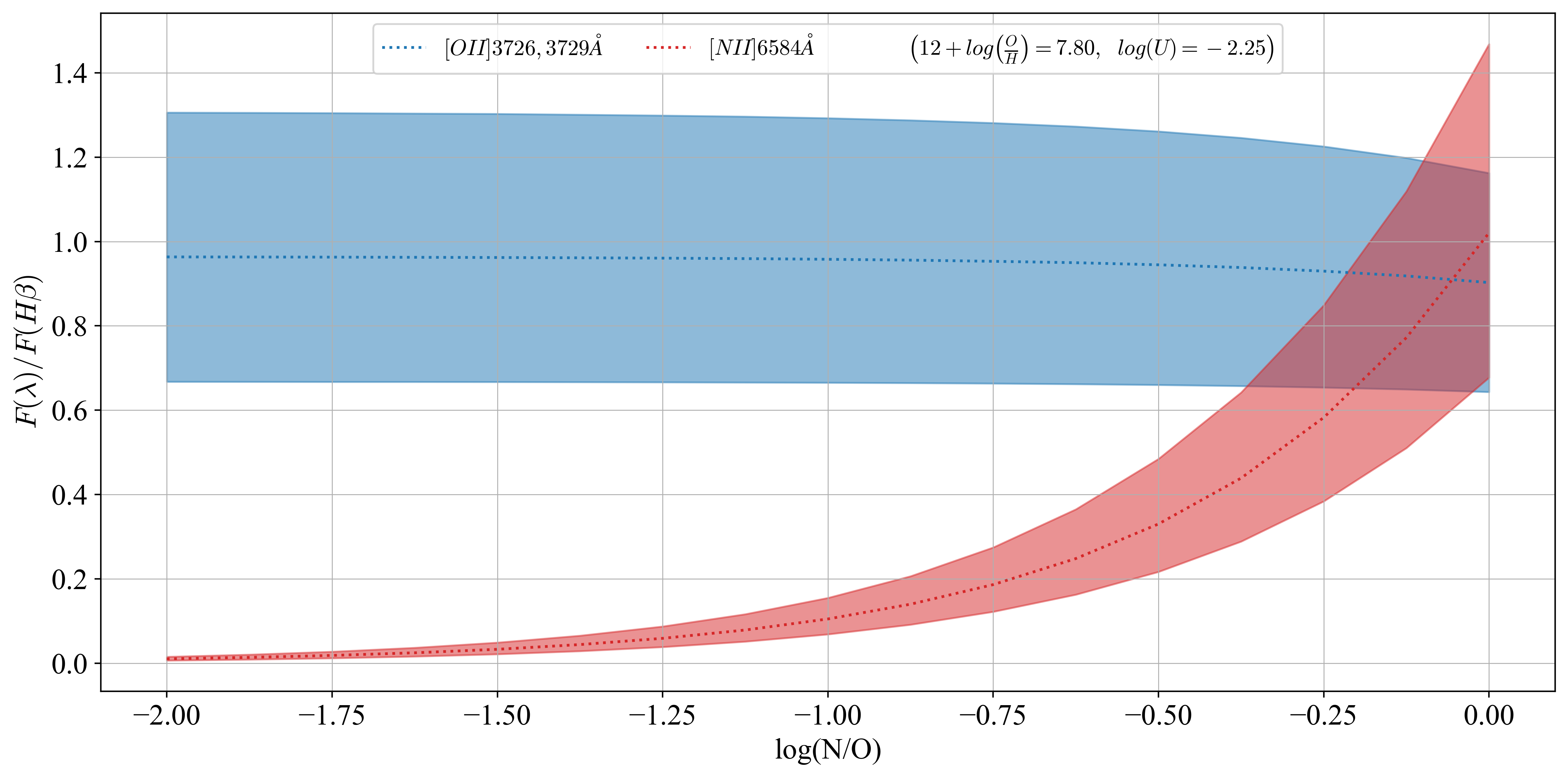}
\includegraphics[width=1.0\columnwidth]{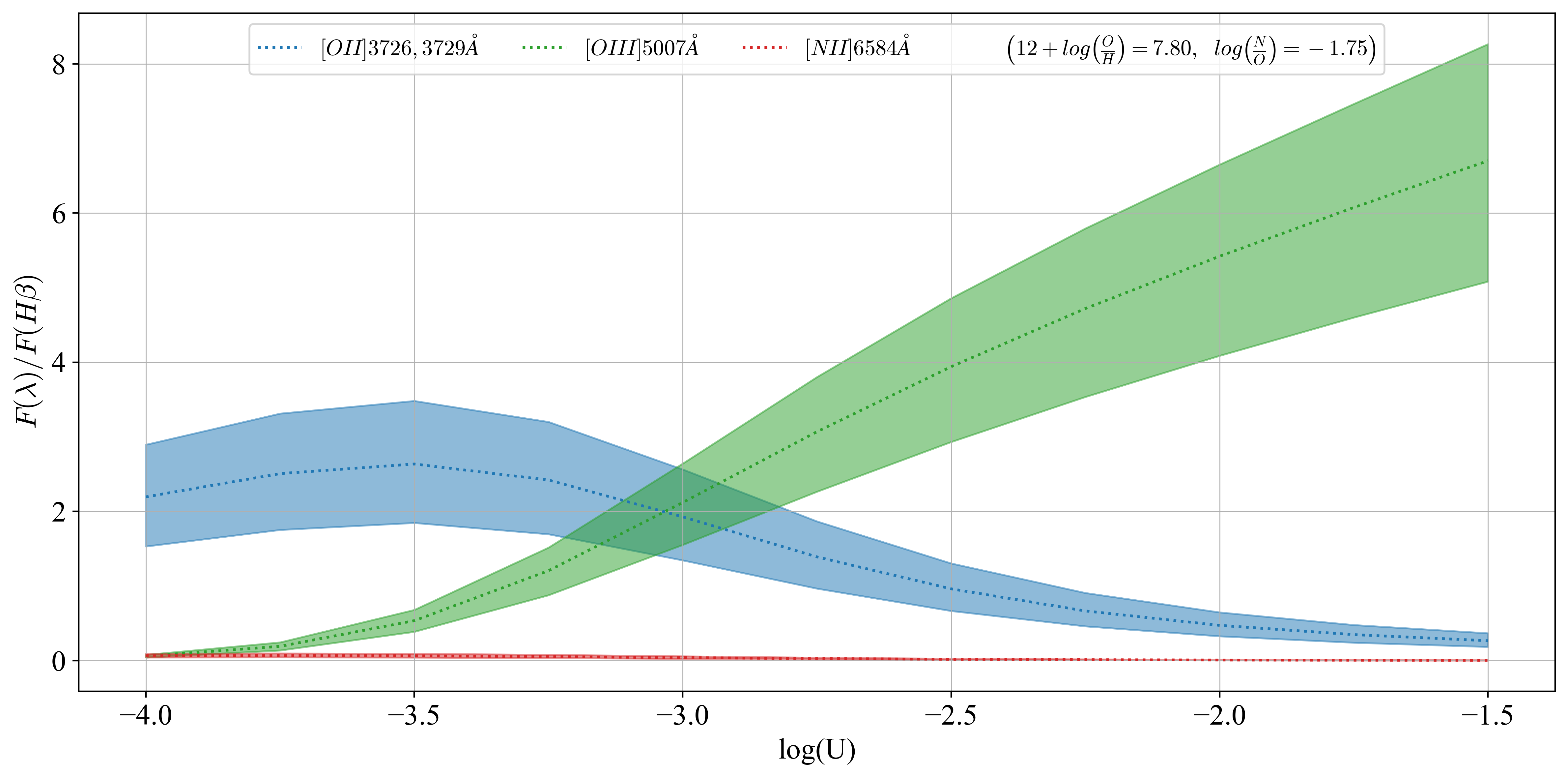}
\caption{\label{fig:convergence_grid} Theoretical fluxes from the photoionization model grids versus the model parameters. a) Evolution of $[OII]3727,3729\AA$ and $[NII]6584\AA$ as a function of the $log(N/O)$ excess. b) Evolution of $[OII]3727,3729\AA$, $[OIII]5007\AA$ and $[NII]6584\AA$ with the ionization parameter. In both plots the dashed lines represent the fluxes for $12+log\left(O/H\right)=7.80$ while the dashed areas cover the $\Delta(O/H)=\pm0.20$ interval in the same scale.}
\end{figure*}

The next step consists in measuring the impact of the error bars on the photoionization model fitting. In Bayesian inference, introducing the measurement error is non-trivial. In this paradigm, the variation on a datum value has nothing to do with observational constraints or instrument limitations. Instead, this dispersion around a mean value is an intrinsic physical property. A major consequence of this interpretation is that the parameters can be modelled by a probabilistic distribution. In \citet{fernandez_bayesian_2019}, the solution proposed was defining a Bayesian likelihood per line, with a normal distribution whose sigma is the line flux uncertainty. Consequently, lines with a larger error have wider likelihoods, which have similar probabilities for a wider range of the parameter values. This is still the approach in the direct method fitting. Moreover, this is the configuration in second row of Table \ref{tab:methods_comparison} labelled as "observed line fluxes and uncertainty" for the photoionization model fitting. The next row contains the results for the "\textsc{HII-CHI-mistry} inputs". In this case, we use the observed fluxes and uncertainty but we only introduced the same lines as in \textsc{HII-CHI-mistry}. In the next test case in Table \ref{tab:methods_comparison} "Uniform maximum observed uncertainty" we overestimate the uncertainty of all the lines to the maximum uncertainty from the input emission spectra. Finally, in the "One order uncertainty range" we overestimate the minimum flux uncertainty one order below the maximum one. For example, if the $[NII]6784\AA$ line has 20\% flux uncertainty, the flux uncertainty of the $[OIII]5007\AA$ flux is overestimated to 2\%.  

The results from the photoionization model fitting and the comparison between the direct method lead to the following insights: 

\begin{itemize}

    \item The measurements from \textsc{HII-CHI-mistry} confirm a hard ionization in CGCG007-025 with a mean ionization parameter of $log(U)=-2.44^{0.17}_{0.14}$ for all the voxels. This value decreases from $-2.14^{0.06}_{0.04}$ at the central knot where the most intense cluster is found to a mean value $-2.47^{0.13}_{0.12}$ including voxels in the north, south and outskirts of the central clump. In Fig. \ref{fig:param_maps}, we can see that there is a small west offset between the central knot and the highest $log(U)$ values. The neural sampler finds values within 0.05 dex with the same spatial behaviour and with slightly higher uncertainty in the measurement. \citet{senchyna_ultraviolet_2017} provided a measurement for this parameter of $log(U)=-4.00\pm0.25$ from the ultraviolet fluxes of CGCG007-025. However, this value was calculated from the photoionization models of \citet{gutkin_modelling_2016} where $log(U)=-4.00$ is the lowest value from the grid. This suggest a typo or issues during the model fitting. Even though there isn't an equivalent for $log(U)$ in the direct method, we have included the $S^{2+}/S^{+}$ map on Fig. \ref{fig:param_maps}. We can observe the same spatial behaviour on this proxy.
    
    \item The mean nitrogen excess reported by \textsc{HII-CHI-mistry} is $log(N/O)=-1.47^{0.03}_{0.04}$. This value is significantly larger than the direct method measurement at $log(N/O)=-1.84^{0.12}_{0.12}$. The results from the neural sampler are similar with slightly lower mean values. Moreover, for all the photoionization model fitting approaches the uncertainty varies very little across the galaxy. This discrepancy can be explained by the degeneracy of the models with the nitrogen excess. This can be appreciated in the left plot of Fig.\ref{fig:convergence_grid}. Looking at the $[NII]6584\AA$ grid as a function of $log(N/O)$, for values $log(N/O)<1.25$ the increase in the relative $[NII]$ photon flux is below 3\% in $7.60<12+log(O/H)<8.00$ range. At this regime the oxygen abundance (which parameterises the metallicity in this model) has a greater impact on the model nitrogen transition flux. Consequently, for a galaxy such as CGCG007-025 with very little $log(N/O)$ excess, anchoring the parameter is challenging. The long-slit observations by \citet{izotov_systematic_2004} reported $log(N/O)_{1}=-1.622\pm0.014$ and $log(N/O)_{2}=-1.622\pm0.014$. In contrast,from the three slits of \citet{van_zee_oxygen_2005} the measurements were  $log(N/O)_{WN}=-1.55\pm0.07$, $log(N/O)_{W}=-1.47\pm0.06$, $log(N/O)_{S}=-1.47\pm0.06$ 
    
    \item The oxygen abundance measured by \textsc{HII-CHI-mistry} shows a large spatial gradient compared with the direct method. At the central knot, the oxygen content is 0.15 dex lower, with a 0.01 dex sigma. In contrast, the outer voxels mean value is 0.1 dex higher, though with a skewed uncertainty at $12+log(O/H)=7.98^{0.04}_{0.11}$ for lower values. This issue can be appreciated on the right hand side plot in Fig.\ref{fig:convergence_grid}. The plot shows a few emission line fluxes evolution with $log(U)$ at $12+log(O/H)=7.80$. In the $-2.5<log(U)<-2.0$ range, $[OIII]$ and $[OII]$ fluxes have a well define behaviour unlike other input fluxes. However, without the $[OII]$ fluxes in the fitting there is a great degeneracy to anchor the $O^{+}$ content. Consequently, at the high ionization voxels, where the $O^{+}$ fraction is expected to be smaller the oxygen abundance is underestimated, while overestimated at lower $log(U)$ values. This can be appreciated in Fig.\ref{fig:maps_methodologies} where the \textsc{HII-CHI-mistry} metallicity decreases in regions as the $log(U)$ increases.
    
    \item In the neural sampler, those configurations with $[OII]7319,7330\AA$ had a better match with the oxygen abundance from the direct method. It should be noted, however, that there isn't an abundance underestimation at the central core unlike in the case of \textsc{HII-CHI-mistry}. This can be explained by the fact that in this techniques all lines are contributing to all the parameters fitting. Consequently, the $[SII]$ and $[NII]$ fluxes are anchoring the  metallicty fraction from the low ionization potential species.
    
    \item In region 2 voxels, where lower signal-to-noise, and hence higher flux uncertainty is observed, all photoionization models displayed a lower parameter uncertainty than the direct method. In the neural sampler, it is possible to compare the fit fluxes against the observational ones as a quality check. It was appreciated that once there is an important uncertainty discrepancy, the algorithm will "neglect" the weaker lines. This means that while $[OIII]$ fluxes have a better match with the models, the weaker lines show discrepancies above 100\%. Using the maximum line uncertainty for all the lines provided a more uniform mismatch between the line fluxes and the models. However, as we can see in Table \ref{tab:methods_comparison}, this increased the discrepancy with the direct method measurement. As a compromise, the last configuration overestimates the minimum flux uncertainty one order below the line with the largest flux error. This provides the best fidelity between the flux across the lines and the measurements with from the direct method. Still, this method has a better precision than the direct method. 
    
\end{itemize}

The previous observations ratify the position of CGCG007-025 as an EELG. Recently, \citet{diaz_use_2022} analysed the composition of high quality observations of diffuse HII regions and HII galaxies. The first group includes star forming bursts in irregular, spiral galaxies, as well as, our own Galaxy. The second group labels blue compact dwarfs, whose luminosity is dominated by the gas emission. The selection criteria is $Eqw(H\beta)>50\AA$, which imposes a maximum stellar cluster age around $~7\,Myr$ \citep[see][]{terlevich_how_2004}. 
The chemical properties of CGCG007-025 lie at the outskirts of the lowest metallicity diffuse HII regions or the typical values of HII galaxies. The oxygen and sulfur median abundance in this family sample were $12+log(O/H)=7.85$ and $12+log(S/H)=6.27$ with ranges of $~7.0-8.5$ and $~5.7-7.25$ respectively. The remaining properties in Table \ref{tab:ne_Te_cHbeta} are also well represented by the HII galaxy family. The $log(S/O)= -1.73\pm0.074$ distribution from all the voxels matches the $-1.9 \lesssim log(S/O) \lesssim -1.3$ regime observed by \citet{diaz_abundance_1991,dors_sulphur_2016,diaz_use_2022} with ratios below solar, ($log(S/O)_{\odot}=1.6$). Finally, the radiation softness parameter introduced by
\citet{vilchez_determination_1988} can be used to map the radiation hardness. This is defined as:

\begin{equation}
\eta=\frac{O^{+}/O^{++}}{S^{+}/S^{++}}
\label{eq:radiation_softness}
\end{equation}

In CGCG007-025, the median value for eq.\ref{eq:radiation_softness} numerator is $\approx2.2$ while its denominator is $\approx4.5$. We find that for both elements, the twice-ionized atoms dominate the abundance. However, we observed that $S^{+}/S^{++} \approx 2\times O^{+}/O^{++}$. This is a common value for low metallicity, high star formation HII galaxies \citep[see][]{diaz_use_2022}. The mean $log(\eta)\approx-0.3$ value is common in HII galaxies \citep[see][]{hagele_temperature_2006,kumari_hardness_2021}.

\section{Conclusions} \label{sec:conclusion}

At $z=0.00469$, the CGCG007-025 - UGC5205 galaxy pair represents a unique system for the study of star formation. While the former is a gas-rich galaxy undergoing a star forming burst, the latter is almost gas depleted. This manuscript is the first study on the pair by the authors, with a focus on the chemical content of CGCG007-025 using an archive MUSE IFU observation. The authors took this opportunity to test novel algorithms for the treatment of large data sets while maintaining a tailored analysis on the emission features. The main takeaways from this work are:

\begin{itemize}

    \item Over 60000 lines were measured in 7774 voxels. These voxels were organized in a set of masks according to the signal-to-noise of $H\alpha$ and our temperature diagnostic line, $[SII]6312\AA$. This was done with \textsc{LiMe}, a python library developed by the authors for the detection, measurement and profile fitting of lines in astronomical spectra. The library is fully documented and this is the beta release for the community. 
    
    \item The MUSE cube was analysed with the stellar synthesis code FADO, which takes into consideration the nebular continua. The absorptions on the HI and HeI transitions were measured to correct the emission features for the extinction and helium abundance calculation. A future manuscript will focus on the stellar populations of the galaxy pair.
    
    \item The extinction was measured across the galaxy by comparing the ratio between the observed hydrogen fluxes and their theoretical emissivities. At the center of the galaxy, where the brightest star formation clump is found, there is a photon drop in $H\alpha$ and $[OIII]5007\AA$. This is likely due to a loss in CCD linearity at the high flux rate from these transitions. However, the extinction could be measured accurately in this region thanks to the Paschen transitions which could be accurately measured at this region. The logarithmic extinction coefficient decreases sharply from $c(H\beta)=0.50\pm0.06$ at the core to the mean value of the galaxy at $c(H\beta)=0.11\pm0.11$. This higher extinction can be explained by the higher density of a very young stellar cluster of $r \approx 48 pc$. 
    
    \item The direct method was applied on 484 voxels with good $[SII]6312\AA$ detection. The methodology consists in a previously published neural network sampler, which fits the entire parameter space simultaneously. The spatial analysis shows normal distributed content for the analysed oxygen, nitrogen, sulfur, helium and argon ionic species. In the case of oxygen, this distribution is $12+log(O/H)=7.88\pm0.11$. Around half this dispersion can be explained by the uncertainty flux measurement. Except for $n_{e}$ and $c(H\beta)$, there is a weak gradient on these abundances. 
    
    \item Since \textsc{HII-CHI-mistry} inbuilt library does not include the $[OII]7319,7330\AA$ lines, the algorithm had issues to anchor the oxygen abundance. The neural sampler managed to reproduce the expected results once these lines were introduced. Both approaches, however, had issues fitting the $log(N/O)$ excess due to the grid degeneracy at this galaxy low $log(N/O)=-1.84\pm0.12$ value. Finally, the models predict a hard ionization field with $log(U)=-2.44^{0.17}_{0.14}$. 
    
    \item Taking advantage of the neural networks speed and efficiency to explore complex parameter spaces, we developed a workflow to test the convergence on the model grid of \textsc{HII-CHI-mistry}. The results were satisfactory and the measurements on CGCG007-025 managed to overcome the issues with the limited inputs of \textsc{HII-CHI-mistry}. Finally, the authors run several Bayesian likelihood configurations to establish the impact of the flux uncertainty on the parameter measurements. The authors recommend to put a limit on the minimum the uncertainty of the fluxes. This way, both auroral and nebular contribute to the sampling and the results are equivalent to those from the direct method. This sigma limit is one order below the uncertainty of the line with the largest uncertainty. 

All these properties place CGCG007-025 in the extreme emission line galaxy family. This makes the galaxy ecosystem an ideal candidate to understand young star formation episodes at low metallicity, which are common at high redshift. Future work will include new observations and methodologies, to explain the phenomena observed.  
    
\end{itemize}

\section*{Acknowledgements}

The authors also wish to thank an anonymous referee, whose
comments helped to improve the clarity of the paper. V.F. acknowledges financial support provided by FONDECYT grant 3200473. 
R.A. acknowledges support from ANID Fondecyt Regular 1202007. 
P.P. thanks Funda\c{c}\~{a}o para a Ci\^{e}ncia e a Tecnologia (FCT) for managing research funds graciously provided to Portugal by the EU.
This work was supported through FCT grants UID/FIS/04434/2019, UIDB/04434/2020, UIDP/04434/2020 and the project "Identifying the Earliest Supermassive Black Holes with ALMA (IdEaS with ALMA)" (PTDC/FIS-AST/29245/2017).  M.G.V.E acknowledges the support of the UK Science and Technology Facilities Council
\section*{Data Availability}

The inclusion of a Data Availability Statement is a requirement for articles published in MNRAS. Data Availability Statements provide a standardised format for readers to understand the availability of data underlying the research results described in the article. The statement may refer to original data generated in the course of the study or to third-party data analysed in the article. The statement should describe and provide means of access, where possible, by linking to the data or providing the required accession numbers for the relevant databases or DOIs.



\bibliographystyle{mnras}
\bibliography{references_zotero.bib}

\appendix

\section{Some extra material}

If you want to present additional material which would interrupt the flow of the main paper,
it can be placed in an Appendix which appears after the list of references.

\begin{table*}
\caption{\label{tab:atomic-data}Atomic data references for the emission lines considered in the chemical analysis.}
\centering{\input{tables/atomic_data_table}}
\end{table*}

\begin{table}
\caption{\label{tab:Priors-and-likelihood} Priors and likelihood distributions
in our model. The term $X^{i+}$ includes all the ionic metal abundances:
$Ar^{2+}$, $Ar^{3+}$, $Cl^{3+}$, $Fe^{3+}$, $O^{+}$, $O^{2+}$, $Ne^{3+}$,
$N^{+}$, $S^{+}$, $S^{2+}$, $y^{+}$ and $y^{2+}$. The helium abundances are defined in logarithmic scale, while the metals are defined using a $12+log\left(X^{i+}\right)$ notation.}
\centering{\input{tables/priors_conf}}
\end{table}


\bsp	
\label{lastpage}
\end{document}

%% file: tables/example_emission_lines.tex
\begin{tabu}{lccc}%
\hline%
Line&$f_{\lambda}$&$F(\lambda)$&$I(\lambda)$\\%
\hline%
$4740\AA\,[ArIV]$&0.03&$10\,\pm1$&$10\,\pm1$\\%
$4861\AA\,HI$&0.00&$1000\,\pm\,4$&$1000\,\pm\,4$\\%
$4922\AA\,HeI$&-0.01&$9\,\pm1$&$8\,\pm1$\\%
$4959\AA\,[OIII]_g$&-0.02&$1986\,\pm\,48$&$1946\,\pm\,47$\\%
$4959\AA\,[OIII]-w1_g$&-0.02&$35\,\pm\,23$&$35\,\pm\,23$\\%
$4987\AA\,[FeIII]$&-0.03&$5\,\pm1$&$5\,\pm1$\\%
$5007\AA\,[OIII]_g$&-0.03&$5542\,\pm\,126$&$5377\,\pm\,123$\\%
$5007\AA\,[OIII]-w1_g$&-0.03&$94\,\pm\,25$&$91\,\pm\,24$\\%
$5016\AA\,HeI_g$&-0.04&$19\,\pm\,3$&$18\,\pm\,3$\\%
$5048\AA\,HeI$&-0.04&$2\,\pm1$&$2\,\pm1$\\%
$5192\AA\,[ArIII]$&-0.07&$4\,\pm1$&$4\,\pm1$\\%
$5272\AA\,[FeIII]$&-0.09&$2\,\pm1$&$2\,\pm1$\\%
$5412\,HeII$&-0.12&$1\,\pm1$&$1\,\pm1$\\%
$5413\AA\,[FeIII]$&-0.12&$1\,\pm1$&$1\,\pm1$\\%
$5518\AA\,[ClIII]$&-0.14&$3\,\pm1$&$3\,\pm1$\\%
$5538\AA\,[ClIII]$&-0.14&$2\,\pm1$&$2\,\pm1$\\%
$5755\AA\,[NII]$&-0.18&$1\,\pm\,1$&$1\,\pm\,1$\\%
$5876\AA\,HeI$&-0.21&$125\,\pm\,1$&$103\,\pm\,2$\\%
$6300\AA\,[OI]$&-0.28&$26\,\pm1$&$20\,\pm1$\\%
$6312\AA\,[SIII]$&-0.28&$19\,\pm1$&$15\,\pm1$\\%
$6347\,[SiII]$&-0.29&$1\,\pm\,1$&$1\,\pm\,1$\\%
$6364\AA\,[OI]$&-0.29&$8\,\pm1$&$6\,\pm1$\\%
$6371\,[SiII]$&-0.29&$1\,\pm1$&$1\,\pm1$\\%
$6548\AA\,[NII]_g$&-0.32&$13\,\pm1$&$10\,\pm1$\\%
$6563\AA\,HI-w1_g$&-0.32&$47\,\pm\,7$&$36\,\pm\,6$\\%
$6563\AA\,HI_g$&-0.32&$3582\,\pm\,65$&$2680\,\pm\,73$\\%
$6584\AA\,[NII]_g$&-0.32&$38\,\pm\,1$&$29\,\pm\,1$\\%
$6678\AA\,HeI$&-0.34&$40\,\pm1$&$29\,\pm1$\\%
$6716\AA\,[SII]$&-0.34&$83\,\pm1$&$61\,\pm1$\\%
$6731\AA\,[SII]$&-0.34&$68\,\pm1$&$50\,\pm1$\\%
$7065\AA\,HeI$&-0.39&$54\,\pm1$&$38\,\pm1$\\%
$7136\AA\,[ArIII]$&-0.40&$73\,\pm\,1$&$51\,\pm\,2$\\%
$7170\AA\,[ArIV]$&-0.40&$1\,\pm1$&$1\,\pm1$\\%
$7281\AA\,HeI$&-0.41&$8\,\pm1$&$6\,\pm1$\\%
$7319\AA\,[OII]$&-0.42&$21\,\pm1$&$14\,\pm1$\\%
$7330\AA\,[OII]$&-0.42&$17\,\pm1$&$12\,\pm1$\\%
$7530\AA\,[ClIV]$&-0.44&$1\,\pm1$&$1\,\pm1$\\%
$7751\AA\,[ArIII]$&-0.47&$18\,\pm1$&$12\,\pm1$\\%
$8045\AA\,[ClIV]$&-0.50&$2\,\pm1$&$1\,\pm1$\\%
$8359\AA\,HI$&-0.53&$5\,\pm1$&$3\,\pm1$\\%
$8374\AA\,HI$&-0.53&$2\,\pm1$&$1\,\pm1$\\%
$8446\AA\,HeI$&-0.53&$10\,\pm1$&$6\,\pm1$\\%
$8467\AA\,HI$&-0.54&$7\,\pm1$&$4\,\pm1$\\%
$8545\AA\,HI$&-0.54&$9\,\pm1$&$6\,\pm1$\\%
$8665\AA\,HI$&-0.55&$13\,\pm1$&$8\,\pm1$\\%
$8750\AA\,HI$&-0.56&$18\,\pm1$&$11\,\pm1$\\%
$8863\AA\,HI$&-0.57&$23\,\pm1$&$14\,\pm1$\\%
$9015\AA\,HI$&-0.58&$33\,\pm1$&$20\,\pm1$\\%
$9069\AA\,[SIII]$&-0.59&$180\,\pm\,2$&$106\,\pm\,4$\\%
\hline%
$c(H\beta)$&&$0.39\,\pm\,0.03$&\\%
$-W(\beta)(\AA)$&&$280\,\pm\,21$&\\%
$F(H\beta) $&&$89\,\pm\,1$\\%
\hline%
\end{tabu}

%% file: tables/absorptions.tex
\begin{tabu}{lcc}%
\hline%
Line&Relative absorption&Voxels\\%
\hline%
$4861\AA\,HI$&1.0&7419\\%
$4922\AA\,HeI$&$0.10 \pm 0.03$&528\\%
$5016\AA\,HeI$&$0.06 \pm 0.02$&879\\%
$5876\AA\,HeI$&$0.05 \pm 0.02$&253\\%
$6563\AA\,HI$&$0.49 \pm 0.22$&7364\\%
$6678\AA\,HeI$&$0.05 \pm 0.01$&48\\%
$8750\AA\,HI$&$0.08 \pm 0.03$&229\\%
$8863\AA\,HI$&$0.10 \pm 0.03$&506\\%
$9015\AA\,HI$&$0.02 \pm 0.02$&708\\%
$9229\AA\,HI$&$0.04 \pm 0.02$&549\\%
\hline%
\end{tabu}

%% file: tables/direct_method_ionic_abundances.tex
\begin{tabu}{lcccc}%
\hline%
Parameter&All voxels&Mask 0 (11 voxels)&Mask 1 (91 voxels)&Mask 2 (382 voxels)\\%
\hline%
$n_{e}(cm^{-3})$&$104.0^{84.0}_{47.0}\,(51.0)$&$378.0^{34.0}_{63.0}\,(63.0)$&$177.0^{84.0}_{64.0}\,(53.0)$&$91.0^{59.0}_{38.0}\,(49.0)$\\%
$T_{low}(K)$&$14776.0^{1322.0}_{1345.0}\,(870.0)$&$14802.0^{204.0}_{169.0}\,(281.0)$&$15030.0^{1192.0}_{643.0}\,(440.0)$&$14577.0^{1527.0}_{1361.0}\,(979.0)$\\%
$c(H\beta)$&$0.21^{0.08}_{0.08}\,(0.02)$&$0.54^{0.06}_{0.04}\,(0.01)$&$0.27^{0.07}_{0.08}\,(0.01)$&$0.19^{0.07}_{0.07}\,(0.02)$\\%
$\frac{Ar^{3+}}{H^{+}}$&$4.91^{0.17}_{0.26}\,(0.13)$&$5.06^{0.01}_{0.04}\,(0.03)$&$4.92^{0.11}_{0.16}\,(0.06)$&$4.87^{0.25}_{0.31}\,(0.17)$\\%
$\frac{Ar^{2+}}{H^{+}}$&$5.27^{0.08}_{0.08}\,(0.05)$&$5.25^{0.02}_{0.04}\,(0.02)$&$5.27^{0.04}_{0.09}\,(0.03)$&$5.27^{0.1}_{0.09}\,(0.06)$\\%
$\frac{O^{+}}{H^{+}}$&$7.36^{0.22}_{0.2}\,(0.12)$&$7.18^{0.04}_{0.03}\,(0.04)$&$7.2^{0.14}_{0.09}\,(0.06)$&$7.43^{0.19}_{0.23}\,(0.14)$\\%
$\frac{O^{2+}}{H^{+}}$&$7.71^{0.09}_{0.11}\,(0.06)$&$7.79^{0.02}_{0.01}\,(0.02)$&$7.74^{0.05}_{0.09}\,(0.03)$&$7.7^{0.1}_{0.11}\,(0.06)$\\%
$\frac{N^{+}}{H^{+}}$&$5.53^{0.14}_{0.14}\,(0.06)$&$5.33^{0.02}_{0.02}\,(0.03)$&$5.4^{0.1}_{0.05}\,(0.03)$&$5.57^{0.12}_{0.14}\,(0.06)$\\%
$y^{+}$&$0.076^{0.004}_{0.003}\,(0.003)$&$0.071^{0.001}_{0.001}\,(0.001)$&$0.076^{0.003}_{0.002}\,(0.002)$&$0.077^{0.004}_{0.004}\,(0.003)$\\%
$\frac{S^{+}}{H^{+}}$&$5.26^{0.15}_{0.15}\,(0.05)$&$5.03^{0.03}_{0.01}\,(0.02)$&$5.13^{0.1}_{0.05}\,(0.02)$&$5.3^{0.12}_{0.13}\,(0.06)$\\%
$\frac{S^{2+}}{H^{+}}$&$5.91^{0.09}_{0.08}\,(0.05)$&$5.88^{0.02}_{0.02}\,(0.02)$&$5.9^{0.04}_{0.07}\,(0.03)$&$5.92^{0.1}_{0.09}\,(0.06)$\\%
\hline%
\end{tabu}

%% file: tables/methodology_results.tex
\begin{tabu}{cccccc}%
\hline%
Methodology&Parameter&All voxels&Region 0 (11 voxels)&Region 1 (91 voxels)&Region 2 (382 voxels)\\%
\hline%
\multirow{7}{*}{Direct method}&$\frac{O}{H}$&$7.88^{0.11}_{0.11}$&$7.88^{0.03}_{0.01}$&$7.87^{0.05}_{0.09}$&$7.88^{0.12}_{0.12}$\\%
&$\frac{N}{O}$&$-1.84^{0.12}_{0.12}$&$-1.85^{0.02}_{0.02}$&$-1.81^{0.08}_{0.07}$&$-1.85^{0.13}_{0.14}$\\%
&$\frac{N}{H}$&$6.04^{0.06}_{0.06}$&$6.03^{0.02}_{0.01}$&$6.05^{0.03}_{0.05}$&$6.03^{0.07}_{0.06}$\\%
&$\frac{Ar}{H}$&$5.43^{0.08}_{0.11}$&$5.46^{0.02}_{0.02}$&$5.43^{0.05}_{0.1}$&$5.43^{0.11}_{0.11}$\\%
&$\frac{S}{H}$&$6.15^{0.1}_{0.09}$&$6.15^{0.02}_{0.01}$&$6.14^{0.05}_{0.08}$&$6.16^{0.11}_{0.1}$\\%
&$ICF(S^{3+})$&$1.47^{0.19}_{0.22}$&$1.66^{0.07}_{0.08}$&$1.48^{0.12}_{0.1}$&$1.42^{0.24}_{0.2}$\\%
&$\frac{S}{O}$&$-1.73^{0.08}_{0.07}$&$-1.73^{0.02}_{0.02}$&$-1.72^{0.03}_{0.03}$&$-1.73^{0.09}_{0.09}$\\%
\hline%
\multirow{3}{*}{$\makecell{\textsc{HII-CHI-mistry}}$}&$\frac{O}{H}$&$7.97^{0.05}_{0.14}$&$7.73^{0.01}_{0.0}$&$7.94^{0.05}_{0.19}$&$7.98^{0.04}_{0.11}$\\%
&$\frac{N}{O}$&$-1.47^{0.03}_{0.04}$&$-1.44^{0.02}_{0.0}$&$-1.46^{0.02}_{0.02}$&$-1.48^{0.04}_{0.04}$\\%
&$log(U)$&$-2.44^{0.17}_{0.14}$&$-2.14^{0.06}_{0.04}$&$-2.28^{0.06}_{0.1}$&$-2.47^{0.13}_{0.12}$\\%
\hline%
\multirow{3}{*}{\makecell{Neural model fitting \\ (observed line fluxes and uncertainty)}}&$\frac{O}{H}$&$7.94^{0.05}_{0.08}$&$7.88^{0.03}_{0.04}$&$7.93^{0.05}_{0.06}$&$7.94^{0.05}_{0.08}$\\%
&$\frac{N}{O}$&$-1.39^{0.04}_{0.04}$&$-1.38^{0.02}_{0.02}$&$-1.4^{0.03}_{0.02}$&$-1.39^{0.05}_{0.05}$\\%
&$log(U)$&$-2.49^{0.15}_{0.17}$&$-2.22^{0.09}_{0.12}$&$-2.39^{0.12}_{0.11}$&$-2.53^{0.14}_{0.16}$\\%
\hline%
\multirow{3}{*}{\makecell{Neural model fitting \\ (same HII-CHI-mistry inputs)}}&$\frac{O}{H}$&$7.96^{0.05}_{0.04}$&$7.92^{0.02}_{0.02}$&$7.95^{0.03}_{0.03}$&$7.97^{0.04}_{0.06}$\\%
&$\frac{N}{O}$&$-1.37^{0.05}_{0.04}$&$-1.3^{0.02}_{0.01}$&$-1.34^{0.03}_{0.03}$&$-1.39^{0.04}_{0.04}$\\%
&$log(U)$&$-2.4^{0.2}_{0.23}$&$-2.04^{0.03}_{0.06}$&$-2.25^{0.14}_{0.14}$&$-2.47^{0.15}_{0.2}$\\%
\hline%
\multirow{3}{*}{\makecell{Neural model fitting \\ (Uniform maximum observed uncertainty)}}&$\frac{O}{H}$&$7.95^{0.09}_{0.08}$&$7.91^{0.02}_{0.0}$&$7.94^{0.06}_{0.03}$&$7.97^{0.08}_{0.1}$\\%
&$\frac{N}{O}$&$-1.33^{0.08}_{0.15}$&$-1.34^{0.02}_{0.02}$&$-1.33^{0.02}_{0.03}$&$-1.33^{0.1}_{0.17}$\\%
&$log(U)$&$-2.3^{0.13}_{0.47}$&$-2.12^{0.02}_{0.03}$&$-2.25^{0.1}_{0.1}$&$-2.33^{0.14}_{0.5}$\\%
\hline%
\multirow{3}{*}{\makecell{Neural model fitting \\ (One order uncertainty range)}}&$\frac{O}{H}$&$7.91^{0.07}_{0.06}$&$7.88^{0.03}_{0.04}$&$7.92^{0.05}_{0.06}$&$7.91^{0.08}_{0.06}$\\%
&$\frac{N}{O}$&$-1.4^{0.05}_{0.05}$&$-1.38^{0.02}_{0.02}$&$-1.4^{0.03}_{0.02}$&$-1.4^{0.05}_{0.06}$\\%
&$log(U)$&$-2.52^{0.17}_{0.19}$&$-2.22^{0.09}_{0.12}$&$-2.39^{0.12}_{0.12}$&$-2.57^{0.17}_{0.16}$\\%
\hline%
\end{tabu}

%% file: tables/atomic_data_table.tex
\begin{tabular}{ccc}
\hline 
Ion & \multicolumn{2}{c}{Atomic data}\tabularnewline
\hline 
\hline 
$H$ & \multicolumn{2}{c}{\citet{storey_recombination_1995}}\tabularnewline
\hline 
$He$ & \multicolumn{2}{c}{\citet{porter_improved_2012}}\tabularnewline
\hline 
$He^{+}$ & \multicolumn{2}{c}{\citet{storey_recombination_1995}}\tabularnewline
\hline 
\hline 
Ion & Collision Strengths & Transition probabilities\tabularnewline
\hline 
\hline 
$O^{+}$ & \citet{pradhan_[o_2006,tayal_oscillator_2007} & \citet{zeippen_transition_1982,wiese_atomic_1996}\tabularnewline
\cline{1-3} 
$S^{+}$ & \citet{tayal_breitpauli_2010} & \citet{podobedova_critically_2009} \tabularnewline
\hline 
$O^{+2}$ & \citet{storey_collision_2014} & \citet{storey_theoretical_2000,wiese_atomic_1996}\tabularnewline
\cline{1-3} 
$N^{+}$ & \citet{tayal_electron_2011} & \citet{wiese_atomic_1996,galavis_atomic_1997}\tabularnewline
\cline{1-3} 
$S^{+2}$ & \citet{hudson_collision_2012} & \citet{podobedova_critically_2009}\tabularnewline
\cline{1-3} 
$S^{+3}$ & \citet{tayal_electron_2000} & \citet{dufton_s_1982,johnson_atomic_1986}\tabularnewline
\cline{1-3} 
$Ar^{+2}$ & \citet{galavis_atomic_1995} & \citet{kaufman_forbidden_1986,galavis_atomic_1995}\tabularnewline
\cline{1-3} 
$Ar^{+3}$ & \citet{ramsbottom_effective_1997} & \citet{mendoza_transition_1982}\tabularnewline
\hline 
\end{tabular}

%% file: tables/priors_conf.tex
\begin{tabular}{cc}
\toprule 
Parameter & Prior distribution\tabularnewline
\midrule
$T_{low}$ & $Normal(\mu=15000\,K,\,\sigma=5000\,K)$\tabularnewline
$T_{high}$ & $Normal(\mu=15000\,K,\,\sigma=5000\,K)$\tabularnewline
$n_{e}$ & $HalfCauchy\left(\mu=2.0,\,\sigma=0\right)$\tabularnewline
$c(H\beta)$ & $HalfCauchy\left(\mu=2.0,\,\sigma=0\right)$\tabularnewline
$X^{i+}$ & $Normal(\mu=5,\,\sigma=5)$\tabularnewline
$y^{+}$ & $Normal\left(\mu=0,\,\sigma=3\right)$\tabularnewline
$y^{2+}$ & $Normal\left(\mu=0,\,\sigma=3\right)$\tabularnewline
$T_{eff}$ & $Uniform\left(min=30000K,\,max=90000K\right)$\tabularnewline
$log(U)$ & $Uniform\left(min=-4.0,\,max=-1.0\right)$\tabularnewline
\midrule 
Parameter & Likelihood distribution\tabularnewline
\midrule
$\frac{F_{X^{i+},\,\lambda}}{F_{H\beta}}$ & $Normal(\mu=\frac{F_{X^{i+},\,\lambda,\,obs}}{F_{H\beta}},\sigma=\frac{\sigma_{X^{i+},\,\lambda,\,obs}}{F_{H\beta}})$\tabularnewline
\bottomrule
\end{tabular}